\documentclass[twocolumn]{aastex631}
\usepackage{amsmath}
\usepackage{comment}

\providecommand{\sorthelp}[1]{}
\defcitealias{2014ApJ...781....5M}{MF14}

\begin{document}

\title{The Atacama Cosmology Telescope: Galactic Dust Structure and the Cosmic PAH Background in Cross-correlation with WISE}

\author[0000-0002-7967-7676]{Rodrigo~C\'ordova~Rosado}
\affiliation{Department of Astrophysical Sciences, Peyton Hall, Princeton University, 4 Ivy Lane, Princeton, NJ, 08544}

\author[0000-0001-7449-4638]{Brandon~S.~Hensley}
\affiliation{Department of Astrophysical Sciences, Peyton Hall, Princeton University, 4 Ivy Lane, Princeton, NJ, 08544}

\author[0000-0002-7633-3376]{Susan~E.~Clark}
\affiliation{Department of Physics, Stanford University, Stanford, CA, 94305}
\affiliation{Kavli Institute for Particle Astrophysics \& Cosmology, Stanford University, PO Box 2450, Stanford, CA, 94305}

\author[0000-0003-2856-2382]{Adriaan~J.~Duivenvoorden}
\affiliation{Center for Computational Astrophysics, Flatiron Institute, 162 5th Avenue, New York, NY 10010 USA}

\author[0000-0002-2287-1603]{Zachary~Atkins}
\affiliation{Joseph Henry Laboratories of Physics, Jadwin Hall, Princeton University, Princeton, NJ, USA 08544}

\author[0000-0001-5210-7625]{Elia~Stefano~Battistelli}
\affiliation{Sapienza University of Rome, Physics Department, Piazzale Aldo Moro 5, 00185 Rome, Italy}

\author[0000-0002-9113-7058]{Steve~K.~Choi}
\affiliation{Department of Physics, Cornell University, Ithaca, NY 14853, USA}
\affiliation{Department of Astronomy, Cornell University, Ithaca, NY 14853, USA}

\author[0000-0002-7450-2586]{Jo~Dunkley}
\affiliation{Joseph Henry Laboratories of Physics, Jadwin Hall, Princeton University, Princeton, NJ, USA 08544}
\affiliation{Department of Astrophysical Sciences, Peyton Hall, Princeton University, 4 Ivy Lane, Princeton, NJ, 08544}

\author[0000-0002-4765-3426]{Carlos~Herv\'ias-Caimapo}
\affiliation{Instituto de Astrof\'isica and Centro de Astro-Ingenier\'ia, Facultad de F\`isica, Pontificia Universidad Cat\'olica de Chile, Av. Vicu\~na Mackenna 4860, 7820436 Macul, Santiago, Chile}

\author[0000-0002-0309-9750]{Zack~Li}
\affiliation{Canadian Institute for Theoretical Astrophysics, University of Toronto, Toronto, ON, Canada M5S 3H8}

\author{Thibaut~Louis}
\affiliation{Universit\'e Paris-Saclay, CNRS/IN2P3, IJCLab, 91405 Orsay, France}

\author[0000-0002-4478-7111]{Sigurd~Naess}
\affiliation{Institute of Theoretical Astrophysics, University of Oslo, Norway}

\author[0000-0002-9828-3525]{Lyman~A.~Page}
\affiliation{Joseph Henry Laboratories of Physics, Jadwin Hall, Princeton University, Princeton, NJ, USA 08544}

\author[0000-0001-6541-9265]{Bruce~Partridge}
\affiliation{Department of Physics and Astronomy, Haverford College, Haverford, PA, USA 19041}

\author[0000-0002-8149-1352]{Crist\'obal~Sif\'on}
\affiliation{Instituto de F\'isica, Pontificia Universidad Cat\'olica de Valpara\'iso, Casilla 4059, Valpara\'iso, Chile}

\author[0000-0002-7020-7301]{Suzanne~T.~Staggs}
\affiliation{Joseph Henry Laboratories of Physics, Jadwin Hall, Princeton University, Princeton, NJ, USA 08544}

\author[0000-0001-5327-1400]{Cristian~Vargas}
\affiliation{Instituto de Astrof\'isica and Centro de Astro-Ingenier\'ia, Facultad de F\`isica, Pontificia Universidad Cat\'olica de Chile, Av. Vicu\~na Mackenna 4860, 7820436 Macul, Santiago, Chile}

\author[0000-0002-7567-4451]{Edward J. Wollack}
\affiliation{NASA/Goddard Space Flight Center, Greenbelt, MD, USA 20771}

\begin{abstract}
We present a cross-correlation analysis between $1\arcmin$ resolution total intensity and polarization observations from the Atacama Cosmology Telescope (ACT) at 150 and 220\,GHz and $15 \arcsec$ mid-infrared photometry from the Wide-field Infrared Survey Explorer (WISE) over 107 $12.5^\circ\times12.5^\circ$ patches of sky. We detect a spatially isotropic signal in the WISE$\times$ACT $TT$ cross power spectrum at $30\sigma$ significance that we interpret as the correlation between the cosmic infrared background at ACT frequencies and polycyclic aromatic hydrocarbon (PAH) emission from galaxies in WISE, i.e., the cosmic PAH background. Within the Milky Way, the Galactic dust $TT$ spectra are generally well-described by power laws in $\ell$ over the range $10^3 < \ell < 10^4$, but there is evidence both for variability in the power law index and for non-power law behavior in some regions. We measure a positive correlation between WISE total intensity and ACT $E$-mode polarization at $1000 < \ell \lesssim 6000$ at $>3\sigma$ in each of 35 distinct $\sim$100\,deg$^2$ regions of the sky, suggesting alignment between Galactic density structures and the local magnetic field persists to sub-parsec physical scales in these regions. The distribution of $TE$ amplitudes in this $\ell$ range across all 107 regions is biased to positive values, while there is no evidence for such a bias in the $TB$ spectra. This work constitutes the highest-$\ell$ measurements of the Galactic dust $TE$ spectrum to date and indicates that cross-correlation with high-resolution mid-infrared measurements of dust emission is a promising tool for constraining the spatial statistics of dust emission at millimeter wavelengths.
\end{abstract}

\section{Introduction} \label{sec:intro}
The interstellar medium (ISM) is a turbulent environment. Energy is injected at large physical scales by processes like stellar feedback, and a complex turbulent energy cascade shapes the ISM over a vast range of physical scales \citep[e.g.,][]{Ferriere:2001, Elmegreen:2004}. High-dynamic range observations of interstellar emission are critical for understanding the flow of mass and energy in the ISM \citep[e.g.,][]{Fissel:2019_decadal, Stinebring:2019}. One common approach is to measure the power spectrum of ISM emission, using tracers like neutral hydrogen (\ion{H}{1}) emission or interstellar dust. These power spectra are often found to be well-described by a power law, with a power spectral index that can be compared to theoretical predictions \citep[e.g.,][]{Crovisier:1983, Miville-Deschenes:2003a, 2005ApJS..157..302M, Miville-Deschenes:2007, Martin:2010, Martin:2015, Blagrave:2017, Pingel:2022}. 

Combining observations of dust emission from Planck and the Wide-field Infrared Survey Explorer \citep[WISE;][]{Wright_2010} with MegaCam measurements of optical scattering from dust, \citet{2016A&A...593A...4M} demonstrated that the dust power spectrum in total intensity (i.e., $TT$) is well-fit by a power law $k^{-2.9\pm0.1}$ from scales of degrees to $\sim1''$, corresponding to physical scales of $\sim0.01$\,pc. The power spectral index of the observable column density is related to the statistics of the 3D density field, which are in turn affected by turbulence and the phase distribution of the gas, but are not directly measurable due to projection effects \citep{Miville-Deschenes:2003b, Clark:2019, Kalberla:2019, 2021ApJ...908..186M}. On scales greater than $5\arcmin$, this index is consistent with measurements across the sky from the Planck satellite \citep{planck2013-pip56, planck2016-XLVIII}. In this work, we investigate the variability of this power law index at smaller scales ($10^3 < \ell < 10^4$, corresponding to $10\arcmin > \theta > 1\arcmin$).

Planck observations established a robust positive correlation between dust total intensity and dust $E$-mode polarization (i.e., $TE$) for multipoles $\ell \lesssim 600$ over much of the sky \citep{planck2016-l11A}. Such a correlation is expected if dust-bearing ISM structures are elongated along magnetic field lines \citep{Zaldarriaga:2001, Huffenberger:2020, Clark:2021}. Indeed, \ion{H}{1} filaments are ubiquitous across the sky with orientations that are strongly correlated with the measured dust polarization angles \citep{Clark_2015, 2019ApJ...887..136C}. \ion{H}{1} structure can thus be used to measure properties of Galactic dust polarization in cross-correlation \citep{Ade:2023, Halal:2023}. Filament-based models successfully reproduce the observed $TE$ correlation \citep{HerviasCaimapo+Huffenberger_2022}. However, in dense regions, it is observed that ISM structures are preferentially oriented perpendicular to magnetic field lines \citep[i.e., negative $TE$;][]{planck2015-XXXV, 2019A&A...632A..17B}. In this work, we extend the characterization of the $TE$ correlation to smaller scales ($1000 < \ell \lesssim 6000$).

To probe the small-scale $TT$ and $TE$ spectra of Galactic dust emission, we employ new maps of millimeter dust emission from the Atacama Cosmology Telescope (ACT) in both total and polarized intensity. The combination of sensitivity, angular resolution, and sky coverage (sky fraction $f_{\rm sky} \simeq 40\%$) afforded by ACT observations enable characterization of dust at arcminute scales over a large sky area, and thus our investigation of the universality of the small-scale dust power spectrum across a range of Galactic environments. We employ ACT maps from two broad bands centered roughly at 150 and 220\,GHz, with angular resolutions of $1.4\arcmin$ and $1.0\arcmin$, respectively.

We complement the ACT data with full-sky observations of mid-infrared emission from WISE. In particular, we use the maps of diffuse emission extracted from observations in the {\it W3} passband by \citet{{2014ApJ...781....5M}} (hereafter \citetalias{2014ApJ...781....5M}). Their custom processing of the WISE data removes compact sources and associated data artifacts. The resulting map covers the full sky at $15''$ resolution with noise properties independent of ACT, enabling correlation analysis down to the ACT resolution limit.

The broad {\it W3} passband is centered at 12\,$\mu$m but has appreciable spectral response from $\simeq$8--16\,$\mu$m \citep{Wright_2010}. For regions typical of the diffuse ISM of the Galaxy, emission at these wavelengths is dominated by the 7.7, 11.3, and 12.7\,$\mu$m mid-infrared emission features associated with polycyclic aromatic hydrocarbons \citep[PAHs; e.g.,][]{Tielens_2008}. Continuum emission is also present from PAHs and likely other nanoparticles undergoing single photon heating. Other expected sources of diffuse emission in this map include the Zodiacal light and the extragalactic background light.

As a ubiquitous component of the Galactic dust population, PAHs are generally well-coupled to the larger grains responsible for far-infrared and millimeter emission \citep[e.g.,][]{Onaka_1996, Mattila_1996, Draine+Li_2007}. Empirically, strong correlations have been observed between the \citetalias{2014ApJ...781....5M} map and dust emission in the Planck bands \citep{Hensley_2016}. However, the mass fraction of dust in the form of PAHs varies over the sky \citep{planck2014-XXIX, Hensley_2016} with an apparent dependence on ISM phase \citep{Hensley_2022}. Further, the PAH emission spectrum is sensitive to the intensity and spectrum of the interstellar radiation field \citep{Draine_2021}. Thus, PAH emission and millimeter dust emission will not be perfectly correlated in detail.

In this work, we assess the ability of high resolution observations of mid-infrared dust emission to correlate with---and thus predict---dust emission properties at millimeter wavelengths. We then use this combination of independent datasets to characterize the Galactic dust power spectrum at small scales both in intensity and polarization, as well as its variation across the sky.

The \citetalias{2014ApJ...781....5M} WISE \textit{W3} map isolates all diffuse emission falling into the {\it W3} passband. In principle, this includes emission from unresolved galaxies across cosmic time. \citet{Chiang+Menard_2019} found that the \citetalias{2014ApJ...781....5M} map has statistically significant correlation with optical measurements of galaxies and active galactic nuclei from redshifts $z \lesssim 2$. They interpret this as redshifted PAH emission from galaxies, i.e., the cosmic PAH background. Indeed, the 7.7\,$\mu$m PAH feature is the strongest of the PAH features and remains in the {\it W3} passband until $z \sim$ 1. In this work, we find robust evidence for a spatially isotropic correlation between the \citetalias{2014ApJ...781....5M} WISE \textit{W3} map and ACT maps at both 150 and 220\,GHz that we interpret as the first detection of the cosmic PAH background in cross-correlation with the Cosmic Infrared Background (CIB).

This paper is organized as follows. In Section~\ref{sec:Data}, we summarize the data products used in this analysis. In Section~\ref{sec:method}, we outline our methodologies for power spectrum computation, uncertainty quantification, and parameter estimation. We present the results of our total intensity $\times$ total intensity (``$TT$'') analysis and our total intensity $\times$ $E$-mode polarization (``$TE$'') analysis in Sections~\ref{sec:TT} and \ref{sec:TE}, respectively. In Section~\ref{sec:diss} we discuss the implications of our results on both the structure of the ISM and on the cosmic PAH background, and we conclude in Section~\ref{sec:conclu}.

\section{Data}\label{sec:Data}

\begin{figure*}
    \centering
    \includegraphics[width=\linewidth]{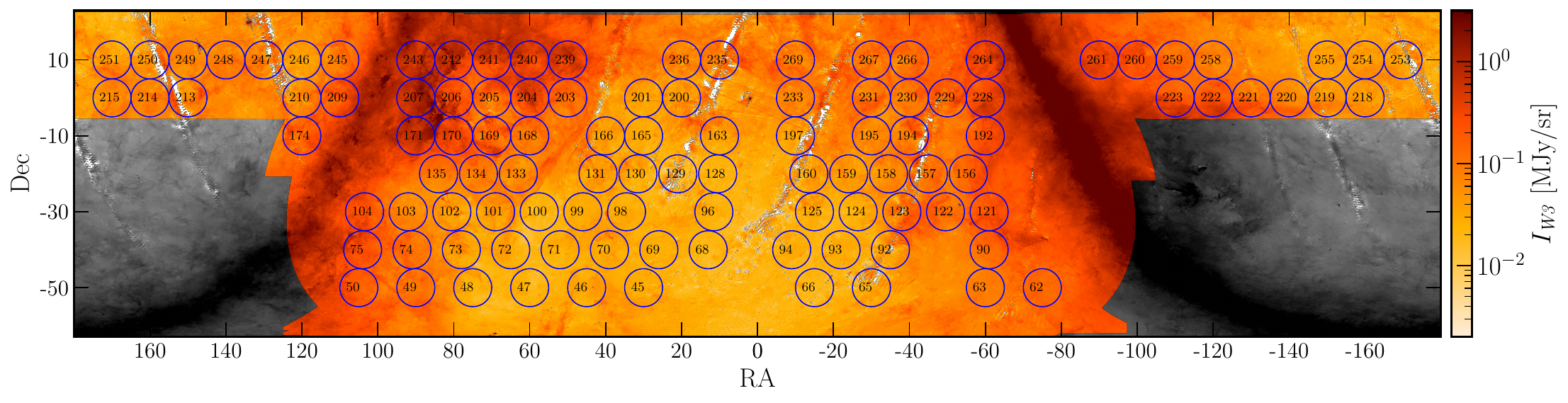}
    \caption{Locations of the 107 \citetalias{2014ApJ...781....5M}  WISE tiles analyzed in this work overlaid on the \citetalias{2014ApJ...781....5M} {\it W3} diffuse emission map. Each $\sim 11^{\circ}$ diameter circle represents the analysis mask adopted in the present study. Areas outside the ACT footprint are rendered in grayscale. White regions, such as the prominent Moon contamination features, are pixels masked by \citetalias{2014ApJ...781....5M}.}
    \label{fig:tiles}
\end{figure*}

\subsection{ACT}\label{sec:ACT}
ACT has measured the total intensity and linear polarization of the millimeter sky over 18,000 deg$^2$. ACT observed in five 
passbands: f030 (22--32\,GHz), f040 (29--48\,GHz), f090 (79--112\,GHz), f150 
(124--171\,GHz) and f220 (171--276\,GHz). In this paper we make use of \textit{I}, \textit{Q}, and \textit{U} Stokes maps made from f150 and f220 observations, with angular resolutions of $1.4'$ and $1.0'$, respectively \citep{2016JLTP..184..772H, 2016ApJS..227...21T,2020JCAP...12..046N}. All of the f150 maps employed in this analysis co-add the three detector arrays that observed in the band; only one array observed in the f220 band.

We use maps made using the f150 and f220 data from the 2008--2019 observing seasons\footnote{These 2008--19 maps are not a major data release for ACT, but the DR5 2008-18 maps are available, and the upcoming DR6 maps from the 2008--22 data will be made public. We compare the measured {\it W3} $\times$ ACT f150 and f220 $TT$ and $TE$ spectra from the 2008--2019 dataset used in this analysis to the spectra computed with the publicly available 2008--2018 DR5 dataset \citep{2020JCAP...12..046N} in two regions. Uncertainties on the power spectra decrease by $\sim$20\% with the inclusion of the 2019 data. A model fit yields parameters that change by less than $1\sigma$ for both $TT$ and $TE$ fits between the datasets.} following a similar process used in making the publicly available Data Release 5 (DR5) maps\footnote{\url{https://lambda.gsfc.nasa.gov/product/act/actpol_prod_table.html}}, which only used data taken through 2018 \citep{2020JCAP...12..046N}. These maps use the same Plate Carrée (CAR) projection as the DR5 maps and have a pixel size of approximately $0\arcmin.5$. These maps do not include any Planck data directly but have been scaled by an overall multiplicative factor that minimizes the residual relative to the Planck temperature power spectrum between $1000 < \ell < 2000$ \citep{2020JCAP...12..046N}. We use maps that include data taken both during the night and day, and from which point sources detected at greater than $5\sigma$ have been subtracted, resulting in a subtraction threshold of $\sim$15~mJy in the f150 band.

We use the ACT-produced ACT+Planck co-added $Q$ and $U$ maps at f150 and f220 for our polarization analysis to enhance our sensitivity at $\ell \lesssim 2000$. The maps are constructed by combining the co-added ACT f150 and f220 maps described above with the Planck PR2 and PR3 143 and 217\,GHz maps \citep{planck2014-a01, planck2016-l01} using the co-adding procedure described in \cite{2020JCAP...12..046N}.

The maps have associated products including inverse variance maps, which include an estimate of the inverse variance in $\mu{\rm K}^{-2}$ per pixel, as well as beam window functions and passbands. The maps are converted from CMB temperature units to ${\rm MJy\,sr^{-1}}$ using conversion factors 395 and 482\,$\rm{MJy~sr^{-1}}~{\rm K}^{-1}$ for f150 and f220, respectively, derived assuming the approximate ACT band centers for a Rayleigh-Jeans spectrum (150 and 220\,GHz, respectively). We consider scales $\ell > 1000$ for the f150 total intensity analysis and all polarization analyses, but only $\ell > 1400$ for the f220 total intensity analysis to avoid scales that are expected to suffer from a loss of power due to modeling errors in the mapmaking process \citep{2022arXiv221002243N}. We use the HEALPix/COSMO polarization convention\footnote{\url{https://lambda.gsfc.nasa.gov/product/about/pol_convention.html}} throughout. The overall systematic uncertainties are at the level of 10\% \citep{2020JCAP...12..046N}, which includes both map calibration uncertainties and unit conversion errors from use of the ACT band centers rather than full bandpass integration.

\subsection{Planck}
We use the Planck 353\,GHz maps\footnote{\url{https://pla.esac.esa.int/\#maps}} produced with the NPIPE data processing pipeline \citep{planck2020-LVII} to extend our analysis to higher frequencies and provide a point of comparison with the ACT measurements. The Planck 353\,GHz channel is the highest Planck frequency designed to measure linear polarization and is the channel most sensitive to polarized thermal dust emission. The maps are provided in units of ${\rm K}$, which we convert to $\rm{MJy~sr^{-1}}$ via a conversion factor of 287.5 $\rm{MJy~sr^{-1}}~{\rm K}^{-1}$ \citep{planck2016-l03}.

\subsection{WISE}\label{sec:WISE}
The Wide-field Infrared Survey Explorer (WISE) satellite observed the full sky in four bands across the mid-infrared \citep[MIR;][]{Wright_2010}. The WISE \textit{W3} band is centered at $\sim$12$\,\micron$ and spans the $8.6$, $11.3$ and $12.7\,\micron$ vibrational emission features of PAHs that dominate Galactic cirrus emission at these frequencies \citep[e.g.,][]{Mattila_1996, Ingalls_2011}.

\citetalias{2014ApJ...781....5M} produced a map of diffuse Galactic emission from the full-sky \textit{W3} data by modeling and subtracting emission from point sources, solar system objects, diffraction spikes, compact sources, and Moon and zodiacal light contamination. \citetalias{2014ApJ...781....5M} combined the \textit{W3} data with Planck 857\,GHz (350\,$\micron$) data to recover extended emission at scales greater than $2^\circ$. The \citetalias{2014ApJ...781....5M} data are provided as a set of 430 $12.5^\circ ~\times~12.5^\circ$ tiles smoothed to an angular resolution of $15\arcsec$ FWHM. Approximately 200 of the 430 tiles overlap significantly ($>50\%$) with the ACT footprint. We smooth these data to a final resolution of $45\arcsec$ before reprojecting onto the ACT pixellization and CAR Projection using the \texttt{pixell}\footnote{Available online at \url{www.github.com/simonsobs/pixell}} package \citep{2021ascl.soft02003N}.

The WISE tiles are provided as fluxes $F_{\it W3}$ in units of digital number (DN), which we convert to MJy\,sr$^{-1}$ following \cite{Cutri_2012}:
\begin{align}
I_\nu &= \frac{F_{\nu 0}}{\theta_{\rm pix}^2} 10^{-\left(M_{0,\rm inst} + \Delta m - 8.926\right)/2.5}\ \frac{F_{\it W3}}{\rm DN} \nonumber \\
&= 0.0135 \frac{F_{\it W3}}{\rm DN}\ {\rm MJy\,sr}^{-1} \label{eq:dn_mjysr}
~~~,
\end{align}
where $F_{\nu0} = 31.674$\,Jy is the {\it W3} zero-magnitude flux density of a source with $F_\nu \propto \nu^{-2}$ \citep{Jarrett_2011}, $\theta_{\rm pix} = 2.75\arcsec$ is the {\it W3} pixel size \citep{Mainzer_2005}, $M_{0, \rm inst} = 17.800$ is the instrumental zero point \citep{Cutri_2012}, $\Delta m = 5.174$ is the conversion from the WISE Vega-system magnitudes to AB magnitudes \citep{Jarrett_2011}, and 8.926 is the factor relating flux density in Janskies to AB magnitudes \citep{Tokunaga_2005}. In detail, the conversion between DN and MJy\,sr$^{-1}$ depends on the spectrum of the source---Equation~\eqref{eq:dn_mjysr} is strictly accurate only for spectra with $F_\nu \propto \nu^{-2}$. However, the conversion factor differs by $\lesssim 13$\% for power law indices between $-3$ and 3 \citep{Wright_2010}.

\section{Methodology}\label{sec:method}

\subsection{Power spectrum estimation} \label{sec:powerspec} 
We compute the angular power spectra in each tile using standard partial sky pseudo-$C_\ell$ methods \citep[e.g.,][]{Hivon_2002}. We compute the $TT$, $TE$, and $TB$ cross-spectra between the WISE \textit{I} total intensity map, and the ACT \textit{I}, \textit{Q}, and \textit{U} maps, using the \texttt{nawrapper}\footnote{Introduced in \citet{actabs2020} and available through GitHub at \url{https://github.com/xzackli/nawrapper}} interface to the \texttt{NaMaster} software  \citep{2019MNRAS.484.4127A}. The mask is described in Section~\ref{sec:masking}. We bin the measurements with $\ell (\ell + 1)$ weighting, with equally-spaced logarithmic bins from $\ell = 1000$ to $10,000$ for f150 and $\ell = 1400$ to $10,000$ for f220 (see Section~\ref{sec:ACT}).

To compute the uncertainty on a binned cross-spectrum $C_b^{xy}$, where $x$ and $y$ are any of $T$, $E$, or $B$, we start from the analytic expression \citep{1995PhRvD..52.4307K, Hivon_2002}

\begin{equation} \label{eq:knox}
    \sigma^2 \left( C_b^{xy}\right) = \frac{(C_b^{xy})^2+(C_b^{xx}C_b^{yy})}{f_{\rm sky} \frac{w_2^2}{w_4}(2 \ell + 1) \Delta \ell}
    ~~~,
\end{equation}
where $f_{\rm sky}$ is the fractional area of the sky, $\Delta \ell$ the bin-width, and $\ell$ the bin midpoint. The correction factor ${w_2^2}/{w_4}$ accounts for the apodization of the mask. The $w_i$ factors are defined as $w_i \equiv \sum \Omega_j W_j^i$, where $\Omega_j$ is the pixel area and $W_j$ is the value of the apodized mask in pixel $j$ \citep{Hivon_2002}. The auto power spectrum term is the sum of a signal and a noise term (i.e., $C_b^{xx} = S_b^{xx} + N_b^{xx}$), while the cross power spectrum is signal-only (i.e., $C_b^{xy} = S_b^{xy}$).

However, Equation~\eqref{eq:knox} neglects the effects of mode-coupling induced by the mask. To account for this, we employ the \texttt{NaMaster} implementation of analytic methods to compute the full covariance matrix \citep{garciagarcia:2019}. Since calculation of the full covariance matrix for all the cross-spectra used in our analysis is computationally expensive, we compute it only for the WISE \textit{W3} $\times$ f150 spectrum in each region in $TT$. We then approximate the per-band uncertainties of each of the other spectra ($TT$ at f220 and both $TE$ and $TB$ at f150, f220, and 353\,GHz) using the simpler expression in Equation~\eqref{eq:knox} rescaled by the ratio between the uncertainty computed from the diagonal of the full covariance matrix and from Equation~\eqref{eq:knox} at f150. This increases the uncertainties by $\sim$10\%, with some scale dependence. Note that only the diagonal entries of the covariance matrix are used for plotting error bars and in parameter estimation.

We apply the same correction derived from the $TT$ spectrum to both the $TE$ and $TB$ uncertainties. Because the same sky mask is used for all frequencies and for all of $TT$, $TE$, and $TB$, this is a good approximation.

To further ensure the robustness of this approach, we exclude regions from our analysis for which the off-diagonal terms of the covariance matrix are large. Specifically, if including the off-diagonal elements of the covariance matrix changes the $\chi^2$ of our best-fit model (see Section~\ref{sec:powerlaw}) by more than 2 for 20 degrees of freedom, then that region is discarded. In practice, this removes regions with particularly complicated masks, e.g., many disconnected regions.

Equation~\eqref{eq:knox} accounts for the contribution of sample variance to the total uncertainty. It would be appropriate to neglect sample variance in reporting the cross-power spectrum of emission in a given region, since the error bars are reflecting only how well the particular spectrum of that region is being measured \citep[see, e.g., the discussion in][]{planck2014-XXX}. In contrast, sample variance should be included when fitting a model for the underlying spectrum from which the observed spectrum is drawn. As such model fitting is a principal focus of our analysis, we include sample variance in all error bars throughout this work. Sample variance is typically $\lesssim 10\%$ of the total uncertainty at all $\ell$.

\subsection{Masking}\label{sec:masking}
Our primary mask for each region is a circle in R.A. and Decl. with a diameter of approximately 11$^\circ$. We apodize this mask with the \verb|C1| cosine taper implemented in \texttt{NaMaster} using an apodization scale of $1^\circ$.

The \citetalias{2014ApJ...781....5M} {\it W3} data include a bitmask of data quality flags. We mask all pixels affected by saturated point sources, first and second latent points of source ghosts, line-like effects, and Moon and solar system object contamination (corresponding to Flags 0, 3, 8, 14, 15, 18, and 20), as well as all pixels without a flux measurement. Point sources brighter than 15\,mJy at f150, predominantly active galactic nuclei \citep[AGN][]{Marsden2014}, are masked with circular holes of radius $5\arcmin$. We apply an additional mask of 14 extended sources identified in ACT maps to mitigate contamination from objects like planetary nebulae and resolved galaxies. This mask was created for the upcoming ACT DR6 power spectrum analysis by visual inspection of the maps after the initial source mask was applied.

We do not mask pixels flagged by \citetalias{2014ApJ...781....5M} as ``compact resolved sources'' (Flag~7). These are mostly galaxies, most of which would not be detected in the ACT maps. We do not wish to remove extragalactic signal, particularly in an inhomogeneous way across the sky, as this would complicate our analysis of extragalactic cross-correlations. On the other hand, it is likely that Flag~7 also identifies some compact resolved Galactic sources that, if retained, would affect our power spectra particularly at high-$\ell$. On balance, we prefer to leave all pixels flagged by Flag~7 in our analysis but note that a careful separation of Galactic versus extragalactic compact resolved sources could improve the analysis presented in this work.

We apodize the source and artifact mask at an apodization scale of $18\arcmin$ using the \verb|C1| cosine taper implemented in \texttt{NaMaster}. The $18\arcmin$ apodization scale follows the nominal choices for the upcoming ACT CMB power spectrum analysis. Our final mask combines the source and artifact mask with the primary circular mask. Each region has a single mask that is used for all analyses at all frequencies.

We select WISE tiles that overlap entirely with the ACT footprint, that retain at least 33\,deg$^2$ of sky area after the mask is applied, and that pass the covariance matrix criterion described in Section~\ref{sec:powerspec}. This results in 107 regions encompassing 6,190 square degrees, or 15\% of the sky, after masking. These regions are illustrated in Figure~\ref{fig:tiles} and span roughly two orders of magnitude in dust column density.

\subsection{Power Law Fitting}\label{sec:powerlaw}
We fit simple power-law models to the $TT$ and $TE$ spectra. Following similar analyses \citep[e.g.,][]{planck2016-l11A}, the Galactic signal at each frequency is modeled as a power law of the form $A_d(\ell/\ell_0) ^ {\alpha_d}$. We expect $\alpha_d \approx -3$ and for $A_d$ to scale with the dust intensity.

For the $TT$ spectra we also include an extragalactic signal that can arise from the cross-correlation of the CIB seen by both WISE and ACT. We use a three-component model containing the Galactic component, a clustered extragalactic component (CIB-C), and a Poisson-like extragalactic component (CIB-P), with total power spectrum given by

\begin{equation}\label{eq:TT3comp}
    C_\ell^{TT} = A_d\,\left(\frac{\ell}{1000} \right) ^{\alpha_d} + A_C\,\left(\frac{\ell}{7000} \right) ^{\alpha_C} + A_P 
    ~~~.
\end{equation}
Here $A_C$ is the amplitude of the clustered component, and $\alpha_C$ is its power law index. $A_P$ is the scale-independent amplitude of the Poisson component. We define $A_{\rm CIB} = A_C + A_P$ as a measure of the total extragalactic power at $\ell = 7000$.

We use the Markov Chain Monte Carlo (MCMC) methods implemented in \texttt{emcee} \citep{emcee} to estimate parameters, using a Gaussian likelihood

\begin{equation} \label{eq:lnlike}
-2\ln {\cal L} = (C_b - w_{b\ell} C_\ell)^TQ^{-1}(C_b - w_{b\ell} C_\ell)
~~~,
\end{equation}
up to an additive constant. Here $C_\ell$ is the model vector, $w_{b\ell}$ are the bandpower window functions that weight the model given the effects of the mode coupling matrix, $C_b$ is the binned data vector, and $Q$ is the diagonal covariance matrix.

We do not expect a significant extragalactic contribution to the $TE$ spectra, so we use a simpler two-parameter Galactic model of the form:

\begin{equation}\label{eq:TEfit}
   C_\ell^{TE} = A_d\,\left(\frac{\ell}{1000} \right) ^{\alpha_d}
   ~~~.
\end{equation}
Since the signal-to-noise ratio of the $TE$ spectra is lower than for $TT$, we fix $\alpha_d=-2.5$ in the baseline analysis following fit values at lower multipoles \citep{planck2016-l11A}.

\section{Total Intensity Analysis}\label{sec:TT}

\begin{figure}
    \centering
    \includegraphics[width = \linewidth]{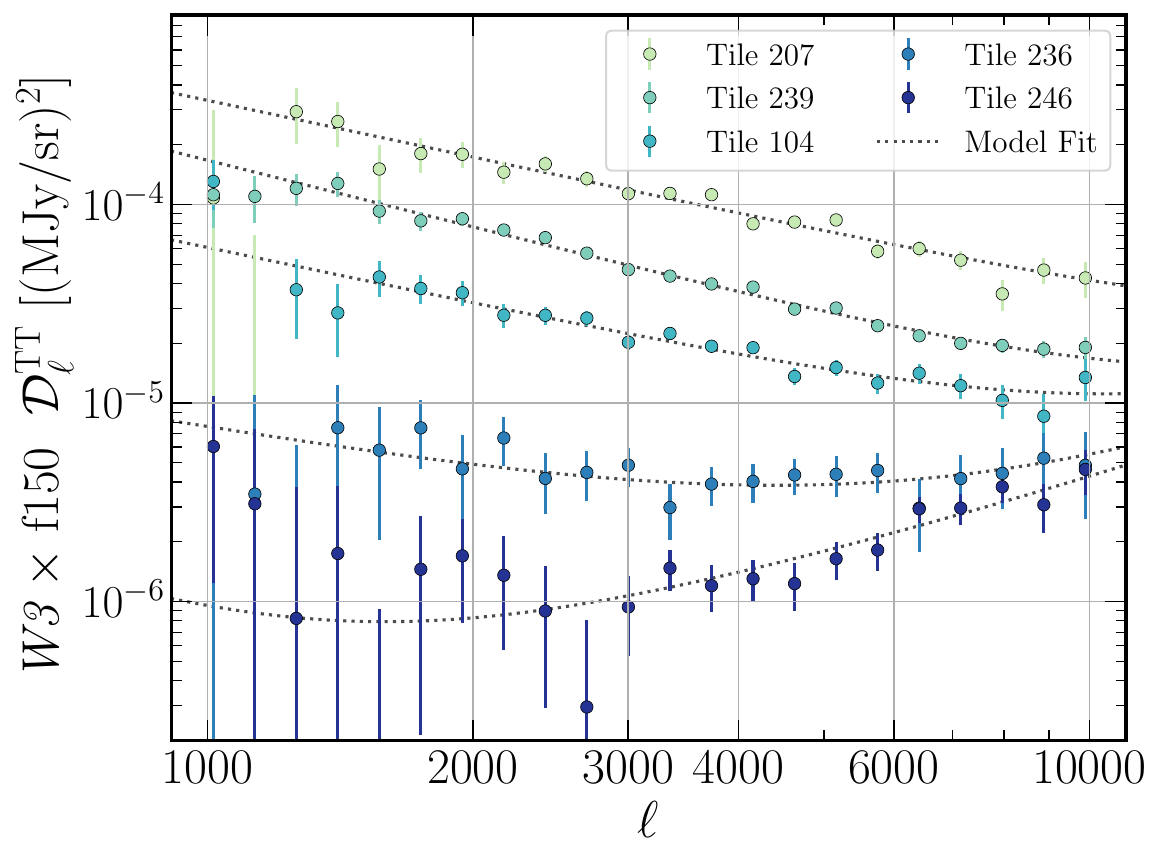}
    \caption{WISE \textit{W3}$\times$ ACT f150 $TT$ spectra of selected regions of varying column density. The slopes of the best fit power laws are consistent with previous measurements of Galactic dust power spectra ($\alpha_d \sim -3$). In lower column density regions we see evidence of an extragalactic component. The best fit model for each region is shown with a dotted line.}
    \label{fig:allTT}
\end{figure}

\begin{figure}
    \centering
    \includegraphics[width=\linewidth]{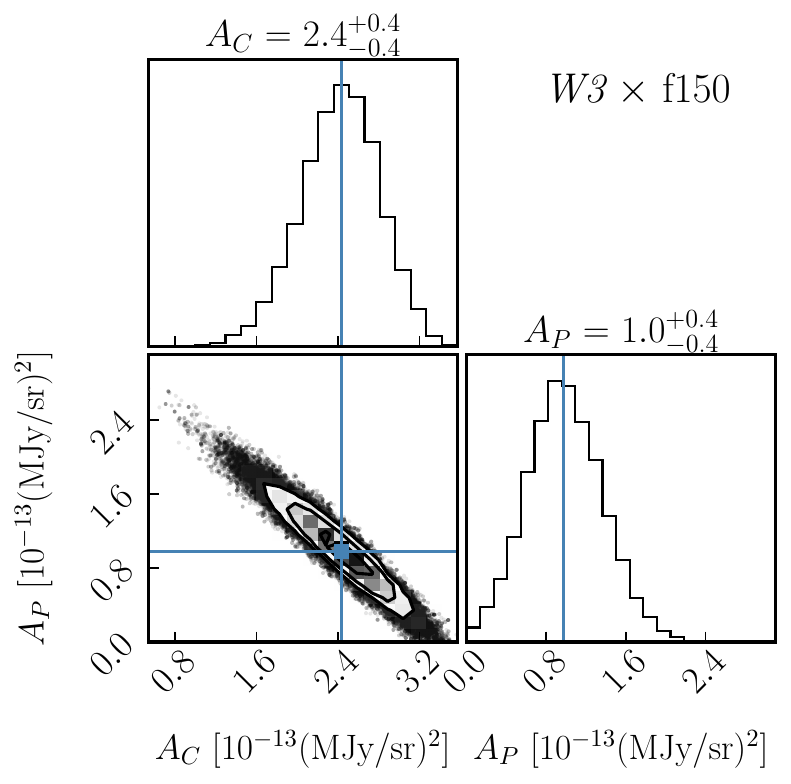}
    \caption{Posteriors for the clustered ($A_C$) and Poisson ($A_P$) components of the fit to the isotropic high-$\ell$ signal in f150 over six low column density regions. We marginalize over the Galactic dust emission in each region by fitting a Galactic dust amplitude ($A_d$) in each region assuming $\alpha_d=-3$. A non-zero signal ($A_C+A_P$) is detected at $30\sigma$ significance.}
    \label{fig:CIB_150MCMC}
\end{figure}

\begin{deluxetable}{LCC}
\tablewidth{0pc}
\tablehead{\colhead{} & \colhead{\textit{W3}$\times$f150} & \colhead{\textit{W3}$\times$f220} \\ \colhead{} & \colhead{[$10^{-12}$ (MJy\,sr$^{-1}$)$^2$]} & \colhead{[$10^{-12}$ (MJy\,sr$^{-1}$)$^2$]}}
\tablecaption{Isotropic Background Parameters \label{table:mcmc_param}} 
\startdata
A_C & 0.24\pm0.04 & 1.2\pm0.2\\  
A_P & 0.10\pm0.04 & 0.2\pm0.2\\ 
A_{\rm CIB} & 0.34\pm0.01 & 1.44\pm0.05
\enddata
\tablecomments{Extragalactic background parameters estimated from WISE $\times$ ACT $TT$ spectra in six regions of low dust column density. A fixed $\alpha_d = -3$ and $\alpha_C = -1$ were assumed (see Equation~\eqref{eq:TT3comp}).}
\end{deluxetable}

\begin{figure*}
    \centering
    \includegraphics[width = \linewidth]{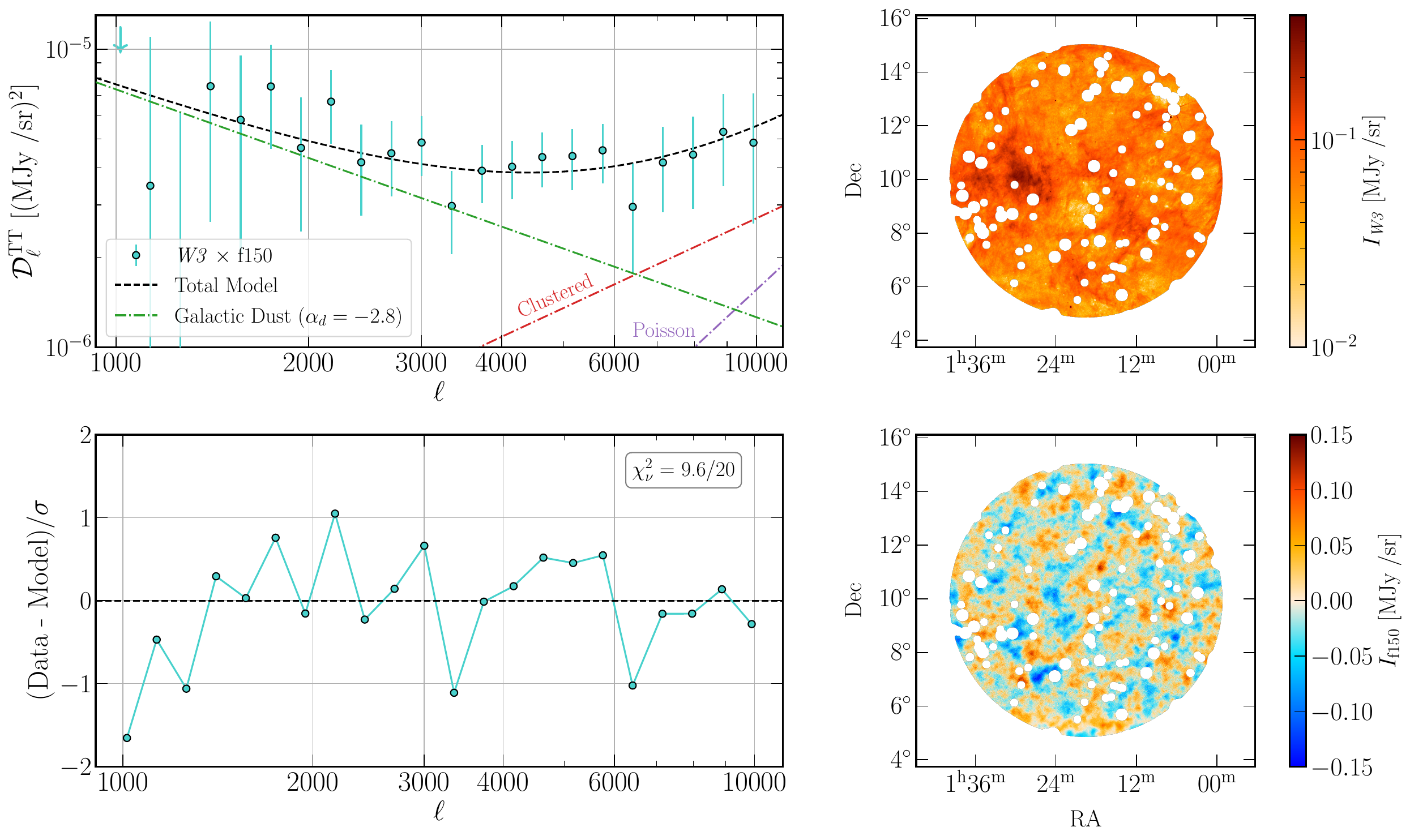}
    \caption{Example model fit of the WISE {\it W3} $\times$ ACT f150 $TT$ spectrum in a moderately high-latitude region ($l = 134.5^\circ, b = -52.2^\circ$; \citetalias{2014ApJ...781....5M} Tile~236). The top left panel shows the measured $TT$ spectrum (blue circles with error bars) where $2\sigma$ upper limits are quoted for bandpowers consistent with zero. Also shown is the total fit model (black) with its region-specific best-fit Galactic dust component (green) along with the global best-fit CIB-C and CIB-P components in red and purple, respectively. The residuals of the fit are presented in the lower left panel. The upper and lower right panels show the WISE and ACT maps of the region, respectively, including the applied mask. The $TT$ spectrum transitions from Galactic emission at low-$\ell$ to extragalactic emission at high-$\ell$ and is well-fit by the model.}\label{fig:236TT}
\end{figure*}

\begin{figure}
    \centering
    \includegraphics[width = \linewidth]{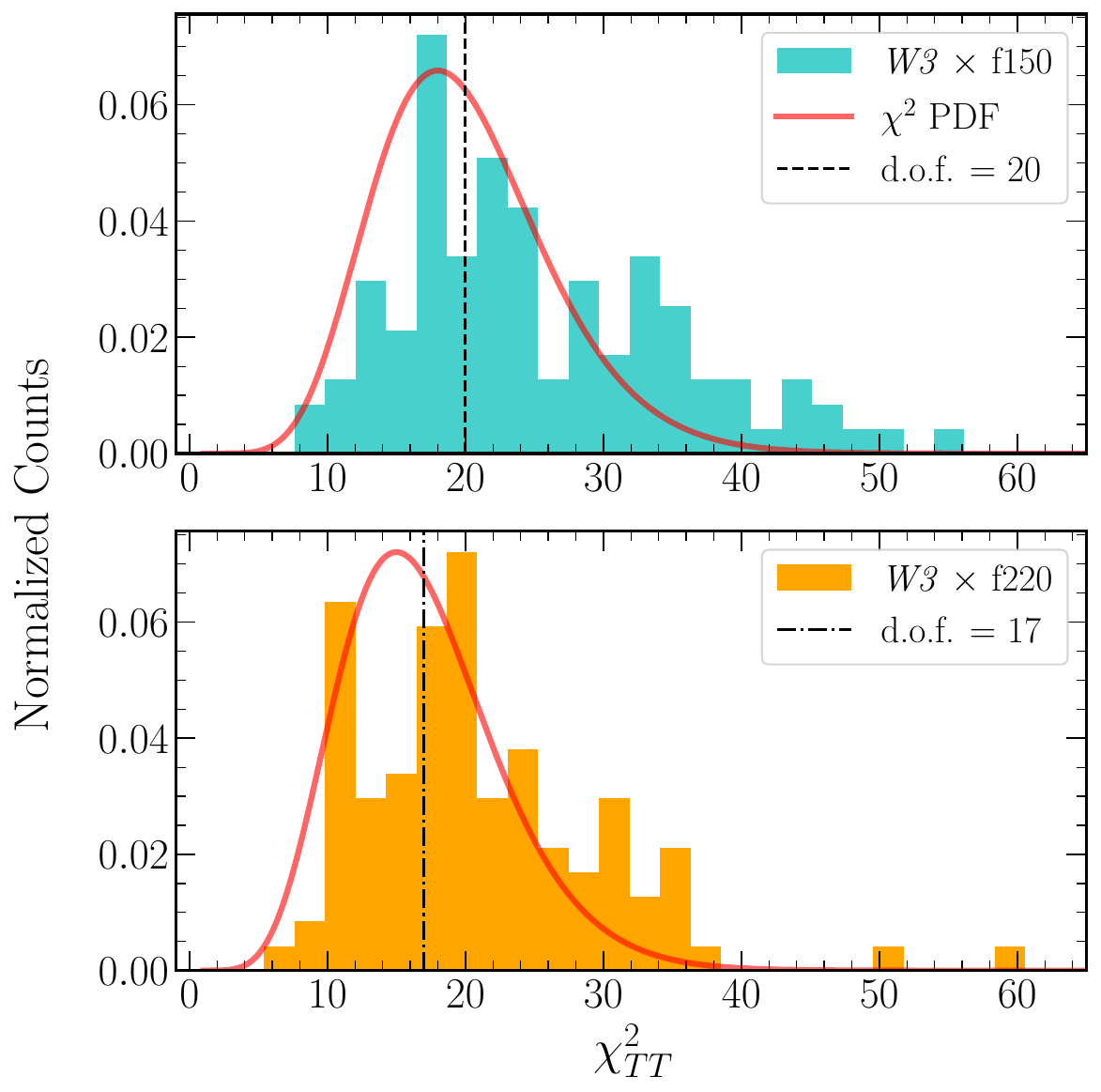}
    \caption{Normalized histograms of $\chi^2$ values for the model fits to all 107 regions at f150 (top) and f220 (bottom). The number of degrees of freedom is indicated with the black dashed line, 20 for f150 and 17 for f220 due to the higher $\ell_{\rm min}$, while the red solid line is the $\chi^2$ probability density function (PDF) for the indicated number of degrees of freedom.} \label{fig:TT_chi2_comparison}
\end{figure}

\begin{figure}
    \centering
    \includegraphics[width = \linewidth]{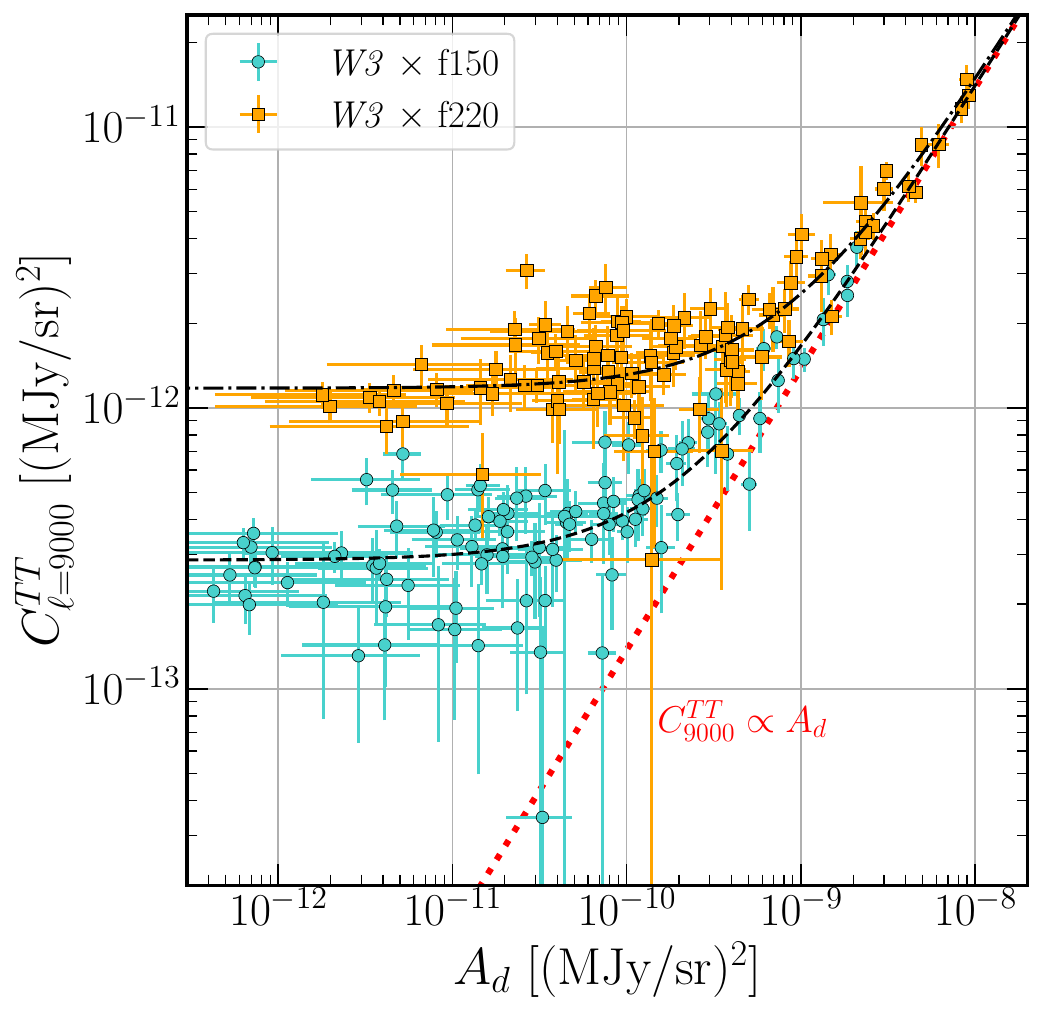}
    \caption{The measured $C_\ell^{TT}$ at $\ell = 9000$ at f150 (teal circles) and f220 (orange squares) in each of the 107 regions is plotted against the fit $A_d$. The red dotted line is $A_d \left(9000/1000\right)^{-3}$, corresponding to the Galactic dust term of Equation~\eqref{eq:TT3comp} for $\alpha_d = -3$. The black lines correspond to Equation~\eqref{eq:TT3comp} with $\alpha_d=-3$ and extragalactic parameters at f150 (dashed) and f220 (dash-dotted) from Table~\ref{table:mcmc_param}. The model provides a good description of the data at both frequencies.} \label{fig:Cell9000}
\end{figure}

\begin{figure}
    \centering
    \includegraphics[width = \linewidth]{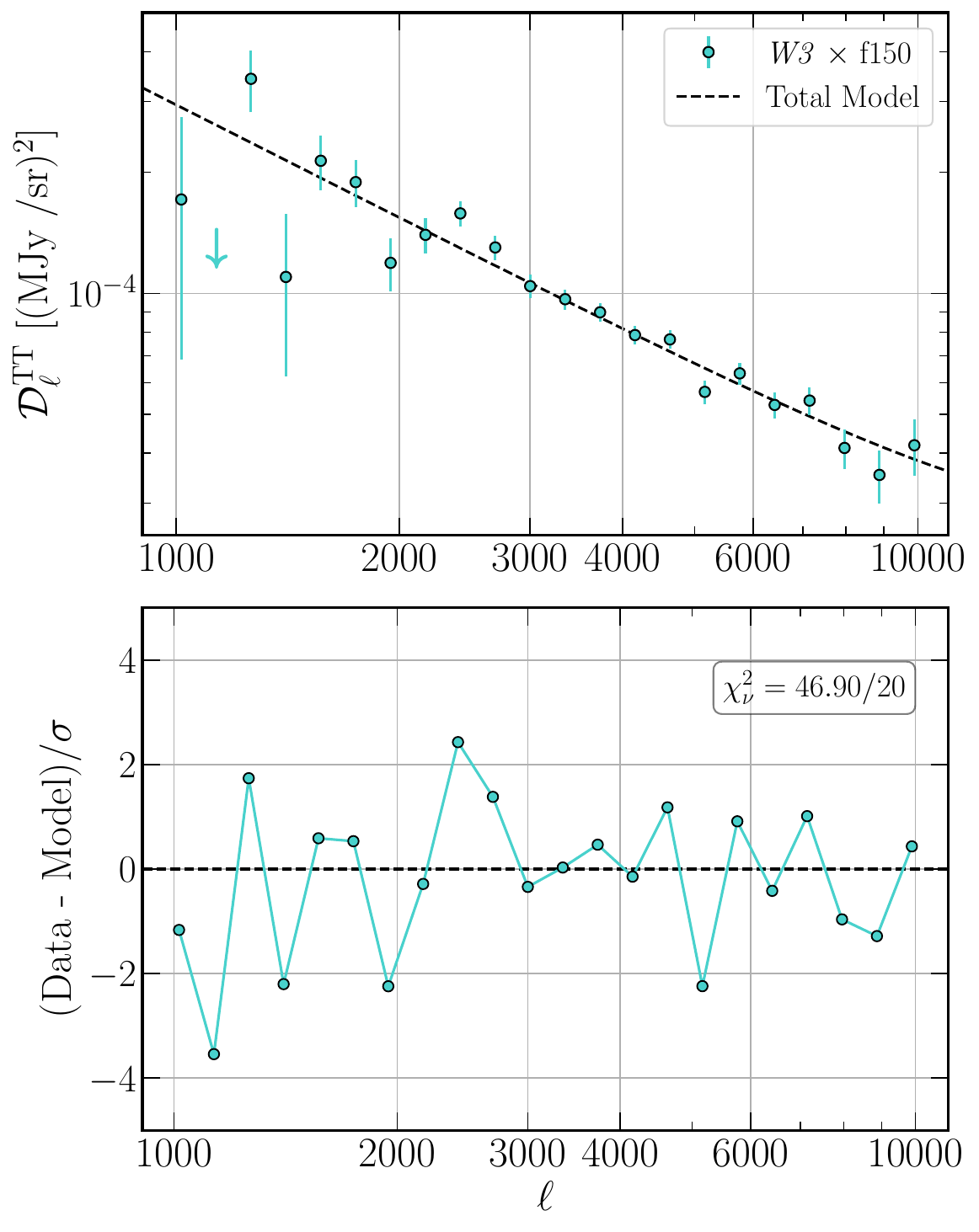}
    \caption{The top panel presents a WISE \textit{W3} $\times$ ACT f150 $TT$ spectrum of a region near the Galactic plane ($\ell = 193^\circ$, $b = -15^\circ$) where the model fit (black dashed) is poor (PTE = 0.06\%). Down arrows indicate 95\% upper limits on bandpowers consistent with zero. The fit residuals are in the bottom panel, demonstrating breakdown of the power law parameterization at $\ell < 3000$.}
    \label{fig:242TT}
\end{figure}

\begin{figure}
    \centering
    \includegraphics[width = \linewidth]{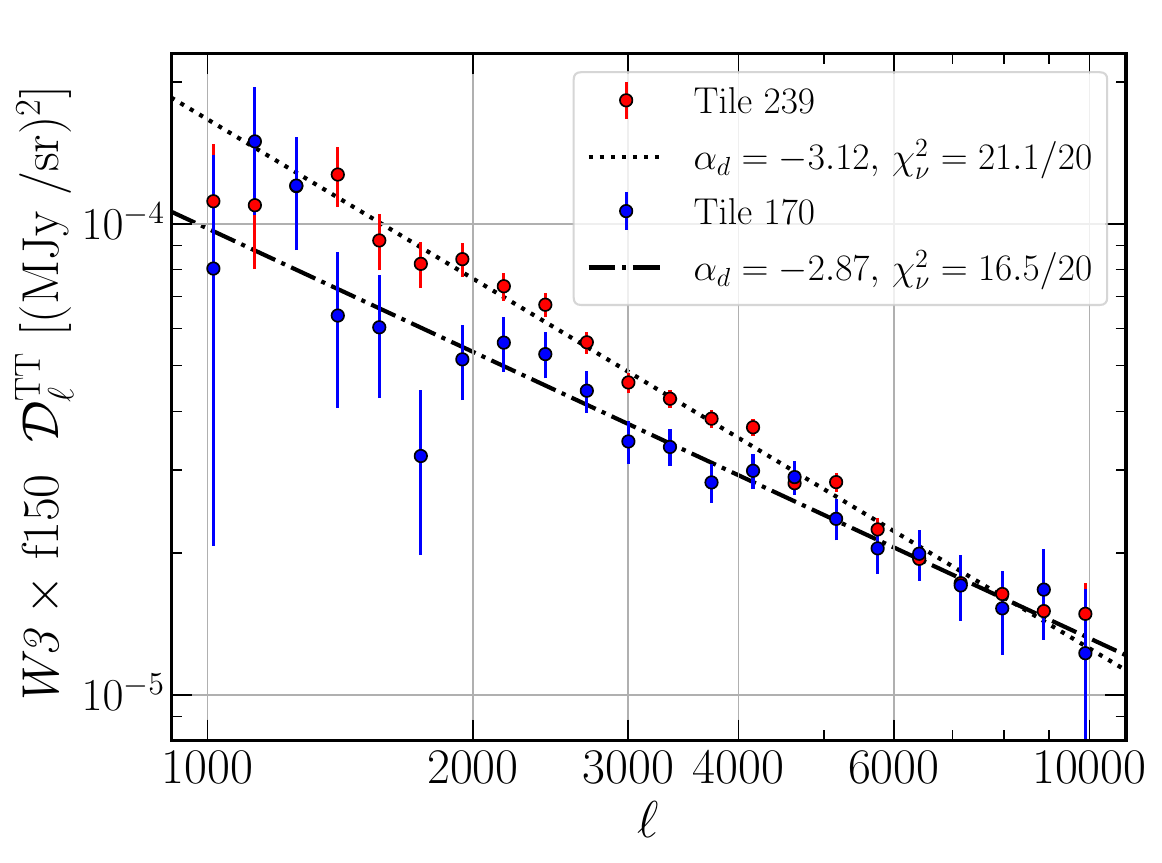}
    \caption{WISE \textit{W3} $\times$ ACT f150 $TT$ spectra in two regions (\citetalias{2014ApJ...781....5M} tiles~170 and 239, located at Galactic latitudes $b = -24.8^\circ$ and $-38.2^\circ$, respectively). The best-fit CIB model has been subtracted from each and the best-fit Galactic dust models are presented as dashed lines. Notably, the measured power law index of the Galactic dust $TT$ spectrum ($\alpha_d$) differs between these tiles at $\sim$3$\sigma$ significance ($-2.87\pm0.07$ versus $-3.12\pm0.03$).}
    \label{fig:TT_var}
\end{figure}

\begin{figure}
    \centering
    \includegraphics[width = \linewidth]{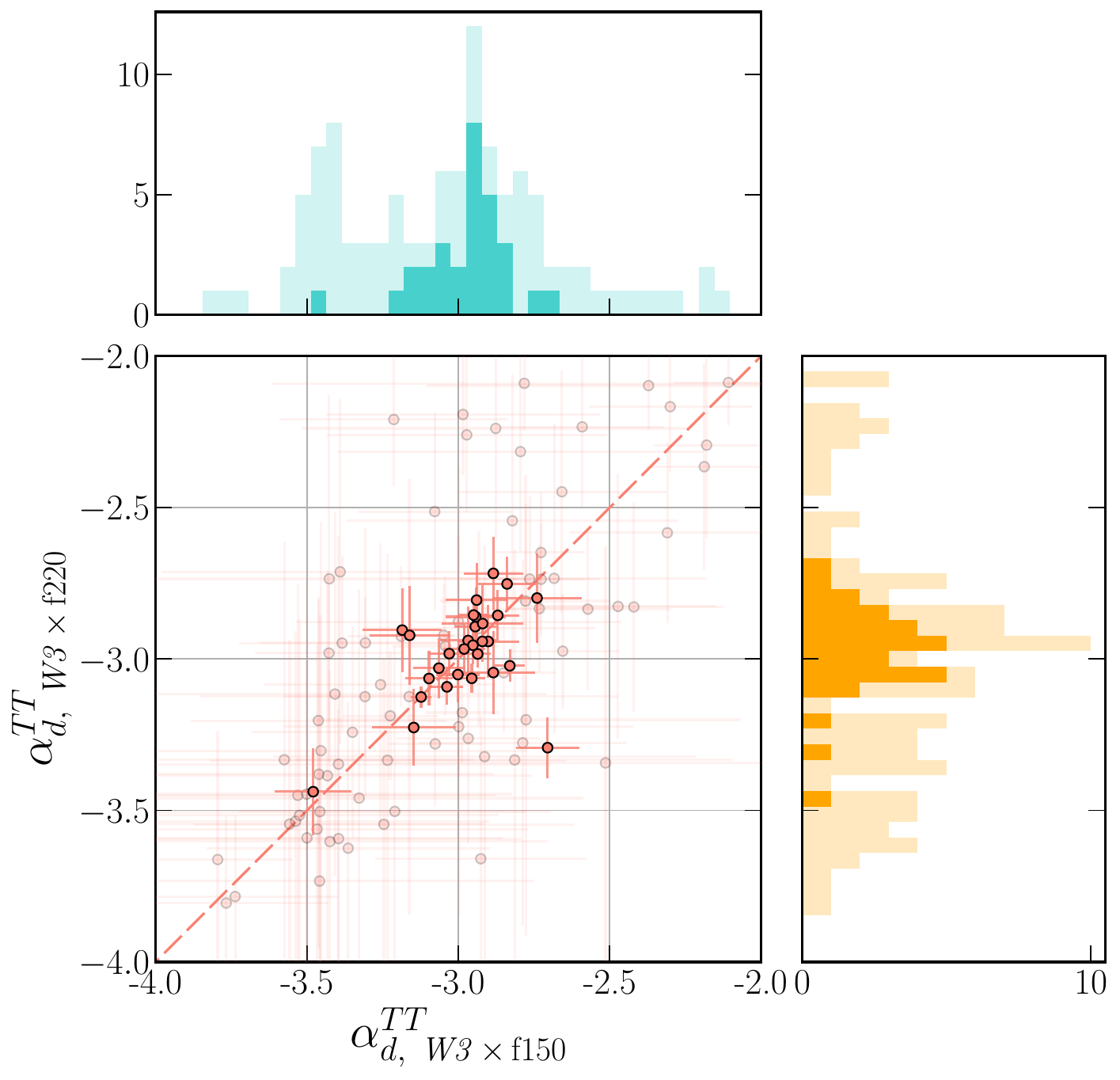}
    \caption{The $\alpha_d^{TT}$ slopes and uncertainties estimated from WISE \textit{W3} $\times$ ACT $[150,\,220]$ GHz. Regions with $> 5\sigma$ detections of Galactic dust $A_d^{TT}$ in both frequencies are opaque. The dashed line shows the one-to-one line. 
    Histograms show the distribution of $\alpha_d$ for the two frequencies, where $> 5\sigma$ measurements of $A_d^{TT}$ are opaque. The dust spectral indices are correlated between f220 and f150.}
    \label{fig:TT_freq_alpha_comp}
\end{figure}

\subsection{Power spectra}
We detect robust correlation between the WISE and ACT maps over nearly the entire region of sky analyzed. Specifically, for 106 of 107 regions in our analysis, we reject the model $C_\ell^{TT} = 0$ at $>3\sigma$ for both {\it W3}$\times$f150 and {\it W3}$\times$f220. A selection of five {\it W3}$\times$f150 $TT$ spectra spanning a range of column densities is presented in Figure~\ref{fig:allTT}. The highest column density tiles shown have roughly power-law spectra with $C_\ell \propto \ell^{-3}$ as has been seen for Galactic dust emission \citep[e.g.,][]{Gautier_1992, Bracco:2011, 2012ApJ...744...40H, planck2013-pip56}. While the lower column density regions of Figure~\ref{fig:allTT} are consistent with this behavior at larger scales ($\ell \lesssim 3000$), all have a rising spectrum in $\mathcal{D}_\ell \equiv \ell(\ell+1)C_\ell/2\pi$ at higher multipoles. Similar behavior is observed with f220 (not shown).

As an initial validation check of the $TT$ spectra, we perform a null test by computing the cross-spectra between WISE data in one region with ACT data in different regions of sky. Specifically, for each of a set of four WISE maps, we compute the {\it W3}$\times$f150 spectrum with ten different ACT regions at the same declination. As expected, all $TT$ spectra are consistent with zero. For the 40 $TT$ spectra consisting of 22 bins each, we find $\chi^2 = 838$ for $C_\ell^{TT} = 0$ versus 880 degrees of freedom (PTE = 84\%).

We therefore seek a physical explanation of the rising $TT$ spectrum at high multipoles. In the following section, we demonstrate that this signal is compatible with extragalactic background fluctuations correlated between WISE and ACT frequencies.

\subsection{Fitting the Extragalactic Background} 
Given that a single power law in $\ell$ is an inadequate description of the $TT$ spectra presented in Figure~\ref{fig:allTT}, we consider the model described in Equation~\eqref{eq:TT3comp} that includes an extragalactic component. Extragalactic emission should have the same amplitude across the sky while the Galactic dust emission should vary from region to region. To avoid assuming a constant Galactic dust $\alpha_d$ across all tiles, we proceed in two steps. First, we infer the extragalactic component parameters $A_C$ and $A_P$ from a simultaneous fit to six regions (Tiles~70, 133, 159, 166, 246, 255) of low column density (median $N_{\rm HI} \lesssim 2.5 \times 10^{20}\,{\rm cm}^{-2}$). We then hold $A_C$ and $A_P$ fixed to their best fit values to fit $A_d$ and $\alpha_d$ in each region separately.

To fit Equation~\eqref{eq:TT3comp} to the six selected regions simultaneously, we first fix $\alpha_d$ and $\alpha_C$ to representative values of $-3$ and $-1$, respectively \citep[see, e.g.,][]{addison:2012}. We then fit the six $A_d$ parameters, one $A_C$ parameter, and one $A_P$ parameter, all with positive definite priors using the methods described in Section~\ref{sec:powerlaw}. The $A_C$ and $A_P$ posteriors for the f150 fit are presented in Figure~\ref{fig:CIB_150MCMC} with best fit values from both the f150 and f220 fits listed in Table~\ref{table:mcmc_param}. We have verified that a simple joint Gaussian likelihood fit to all 107 tiles simultaneously with $\alpha_d = -3$ and $\alpha_C = -1$ yields consistent best-fit values of $A_C$ and $A_P$.

We find that $A_{\rm CIB} = A_C + A_P > 0$ at 30$\sigma$ significance at both f150 ($A_{\rm CIB} = 3.4\pm 0.1 \times 10^{-13}\,{\rm (MJy~sr^{-1})^2}$) and f220 ($A_{\rm CIB} = 1.44\pm0.05 \times 10^{-12}\,{\rm (MJy~sr^{-1})^2}$). The fits yield a ratio $\mathcal{F}_{iso} \equiv A_{\rm CIB}^{\rm f220}/A_{\rm CIB}^{\rm f150} = 4.2\pm0.2$. The systematic uncertainty on these numbers is of order 10\% (see Section~\ref{sec:ACT}). \citet{2013JCAP...07..025D} performed multi-frequency fits to the CIB signal in ACT f150 and f220 using a modified blackbody model $I_\nu \propto \nu^\beta B_\nu(T_d)$ for the frequency dependence, where $B_\nu(T)$ is the Planck function and $T_d$ is the dust temperature. They found $\beta = 2.2\pm0.1$ for a fixed dust temperature $T_d = 9.7\,{\rm K}$, equivalent to $\mathcal{F}_{iso} = 4.1\pm 0.2$. Thus, our derived value is consistent with previous ACT measurements of the CIB at millimeter wavelengths. We have likewise verified that the fitted Galactic dust $A_d$ parameters at f150 and f220 are consistent with a typical frequency scaling for Galactic dust emission, although, by design, the Galactic dust signal is weak in these regions and the constraints are not stringent.

The amplitudes $A_C$ and $A_P$ are strongly anti-correlated, and the Poisson component is not measured at high significance. To investigate this further, we perform a fit with just one extragalactic component where we fit for both amplitude and slope. For the latter, we impose a uniform prior $[-2, 1]$. This fit excludes $\alpha = 0$ at $>5\sigma$, i.e., a pure Poisson component is strongly disfavored. On the other hand, if we fix $\alpha_C = -1$ and $A_P = 0$, we find little degradation in the goodness of fit (PTE = 0.42 versus 0.51 for the fiducial model). In another variation, we fit $A_P$, $A_C$, and $\alpha_C$, imposing a uniform prior on $\alpha_C$ of $[-1.5, -0.5]$. We find that $\alpha_C \leq -0.76$ at 95\% confidence. The data therefore require a component that resembles the clustered component of the CIB but do not require a Poisson component. However, given the extent of the degeneracy between $A_C$ and $A_P$ (see Figure~\ref{fig:CIB_150MCMC}), we cannot place strong constraints on their relative amplitudes.

We have found a high-$\ell$ $TT$ correlation between the WISE and ACT maps that is well-fit with a single amplitude across six regions and with amplitudes at f150 and f220 consistent with the frequency scaling of the CIB. We therefore interpret this signal as correlation between galaxies observed by both WISE and ACT. Possible origins of the extragalactic component are discussed further in Section~\ref{sec:cib}, but we will first verify that this component is indeed of constant amplitude across the remaining 101 regions.

For all subsequent $TT$ fits, we fix $A_C$ and $A_P$ parameters to their fit values at a given frequency (see Table~\ref{table:mcmc_param}) and fix $\alpha_C = -1$.

\subsection{Fitting the Galactic Dust TT Spectrum}

\subsubsection{Goodness of Fit}
In the previous section, we derived the best fit values of the extragalactic parameters in our $TT$ model ($A_C$ and $A_P$ in Equation~\eqref{eq:TT3comp}) based on a set of six regions. In this section, we fix $A_C$ and $A_P$ to these values (see Table~\ref{table:mcmc_param}) and perform another MCMC fit separately in each of the 107 regions to derive $A_d$ and $\alpha_d$. In these fits, $A_d$ is required to be positive and a conservative uniform prior of $[-4,-2]$ is imposed on $\alpha_d$ based on measurements in the literature \citep{2016A&A...593A...4M, planck2016-l11A}.

An example fit to a WISE {\it W3} $\times$ ACT f150 $TT$ spectrum is presented in Figure~\ref{fig:236TT}. This region, Tile~236, is centered on Galactic longitude $l = 134.5^\circ$ and Galactic latitude $b = -52.2^\circ$ and transitions from being dominated by a Galactic dust spectrum at low-$\ell$ to an extragalactic spectrum at high-$\ell$. The Galactic component is best fit with $\alpha_d = -2.8 \pm 0.2$, and overall the parametric model provides an excellent fit to the data ($\chi^2 = 9.6$ for 20 d.o.f). Overall, the model provides a good description of the data in all regions. As illustrated in Figure~\ref{fig:TT_chi2_comparison}, the distribution of $\chi^2$ values across all 107 regions at both f150 and f220 is broadly consistent with expectations, though some outliers have high $\chi^2$ values.

Another visualization of the model fit to all 107 regions is presented in Figure~\ref{fig:Cell9000}, which plots the best-fit Galactic dust amplitude $A_d$ (see Equation~\eqref{eq:TT3comp}) against the measured $\ell = 9000$ bandpower. In high column density regions (large $A_d$), the high-$\ell$ spectrum is dominated by Galactic dust emission and there is a strong linear correlation between $C_{\ell = 9000}^{TT}$ and $A_d$. As $A_d$ decreases to lower column densities, however, $C_{\ell = 9000}^{TT}$ asymptotes to a roughly constant value in both f150 and f220. This is the extragalactic signal common to all regions. The sum of the Galactic and extragalactic model components, plotted in black lines assuming $\alpha_d = -3$, provides a good description of the measurements.

Although the model fits are generally good, Figure~\ref{fig:TT_chi2_comparison} demonstrates that the distribution of $\chi^2$ values is biased toward higher values than expected. Twelve regions in f150 and nine in f220 have fits with PTE $< 1$\%. We identify two possible explanations for the model failures: (1) Galactic dust $TT$ spectra that differ from a pure power law and (2) unmasked compact Galactic sources. Of the 38 (31) regions with PTE $<\,10\%$ ($\chi^2 > 25 $) in f150 (f220), 18 (14) are regions where departures from the model are mostly at $\ell < 3000$, while the remaining 20 (17) are mostly at $\ell > 3000$.

We illustrate an example of (1) in Figure~\ref{fig:242TT}, which presents the {\it W3} $\times$ f150 spectrum of a region near the Galactic plane ($b = -15^\circ$). While the $\ell > 3000$ spectrum is well-fit by a power law ($\alpha_d = -2.93 \pm 0.04$), there are clear departures at lower multipoles. Indeed, the PTE of the fit is only 0.06\%. The relatively high column density of the region permits high signal-to-noise measurements even at $\ell = 10^4$, and thus deviations from our simple parametric model are easier to detect. Further, this region includes a range of dense, complex Galactic structure likely at different distances, and so it is not unexpected that the spatial statistics are complicated. Thus, at least in some regions, we appear to be seeing the inability of the model to capture the complexity of the dust emission. We discuss the implications of this further in Section~\ref{sec:discuss_var}.

While (1) represents a limitation of the model, (2)---the presence of Galactic sources in the maps---is a limitation of the analysis. Sources such as stars, planetary nebulae, and supernova remnants have been identified in ACT maps \citep{2020JCAP...12..046N} and many of these have counterparts in the WISE maps. While we have used maps that are as cleaned of these objects as possible (see Section~\ref{sec:Data}), residual correlations from objects below our flux cuts could contribute power particularly at high-$\ell$. Given the robustness of extragalactic background model fit across a wide range of column densities and Galactic latitudes (see Figure~\ref{fig:Cell9000}), it is unlikely that such sources constitute much of the signal we have identified as extragalactic. However, they could account for regions with high-$\ell$ power in excess of our model. Higher-fidelity modeling of diffuse dust emission will almost certainly require a dedicated effort to identify and mask Galactic sources at lower flux thresholds than employed here.

\subsubsection{Variation in Galactic dust TT spectrum} \label{sec:variability}

We have demonstrated that the model of Equation~\eqref{eq:TT3comp} provides a good description of the $TT$ spectrum of most of the 107 regions analyzed here. Although we fit for $\alpha_d$ in each region individually, Figure~\ref{fig:Cell9000} illustrates that a constant $\alpha_d = -3$ yields a reasonable fit to the data. In this section, we demonstrate that there is true $\alpha_d$ variability in our sample, justifying our choice of fitting $\alpha_d$ as free parameter and having implications for modeling the dust $TT$ spectrum more broadly.

Figure~\ref{fig:TT_var} illustrates an example of variation in the slope of the {\it W3} $\times$ f150 dust $TT$ spectrum between two regions. \citetalias{2014ApJ...781....5M} tiles~170 and 239, both at moderate Galactic latitudes ($b = -24.8^\circ$ and $-38.2^\circ$, respectively), have comparable $TT$ power at $\ell \gtrsim 6000$. However, at lower multipoles they diverge. The best fit $\alpha_d$ values for the two regions are $-3.12\pm0.03$ and $-2.87\pm0.07$, respectively. The {\it W3} $\times$ f220 $TT$ spectra of these two regions are best fit by $\alpha_d$ values of $-3.13\pm0.04$ and $-2.86\pm0.08$, respectively.

The distribution of best fit $\alpha_d$ values is presented in Figure~\ref{fig:TT_freq_alpha_comp}, which shows the best fit $\alpha_d$ in f150 against the best fit value at f220 for 29 regions with $A_d/\sigma(A_d) > 5$ at both frequencies. The best fit values for $\alpha_d$ range from $-3.4$ to $-2.7$. In these 29 regions, we find a median $\alpha_{d} = -2.95$ and $-2.96$ at f150 and f220, respectively, while the $\alpha_d$ values at the two frequencies are correlated at a level of Pearson $r = 0.5$. To assess the statistical significance of region-to-region variation in $\alpha_d$, we consider two models: (1) $\alpha_d$ is constant over all regions, and is estimated as the inverse variance weighted mean of the 58 $\alpha_d$ values in the 29 regions at the two frequencies; and (2) $\alpha_d$ differs from region to region, and is estimated in each region as the weighted mean of the two $\alpha_d$ fits at f150 and f220. Model (1) has $\chi^2 = 151$ for $57$ degrees of freedom while Model (2) has $\chi^2 = 37$ for $29$ degrees of freedom. The likelihood ratio test assuming the $\alpha_d$ posteriors are Gaussian yields a $7\sigma$ preference for the model with region-to-region $\alpha_d$ variations. The model with constant $\alpha_d$ has a PTE of $2\times10^{-10}$, corresponding to exclusion at $6\sigma$ significance.

The correlation of the fit $\alpha_d$ values between f150 and f220 is consistent with true astrophysical variations in the value of $\alpha_d$. However, some of the variation could be driven by fitting degeneracy between the amplitudes of the Galactic component and the extragalactic signal. We find that $\alpha_d$ is negatively correlated with column density, i.e., with shallower slopes at low column densities. This behavior is as predicted by fitting degeneracy but is not inconsistent with astrophysical variation. The $5\sigma$ cut employed above mitigates the effect of this fitting degeneracy on our analysis of $\alpha_d$ variations. If these are indeed physical variations, they could be confirmed with other tracers.

\section{Polarization Analysis} \label{sec:TE}

\begin{figure*}
    \centering
    \includegraphics[width = \textwidth]{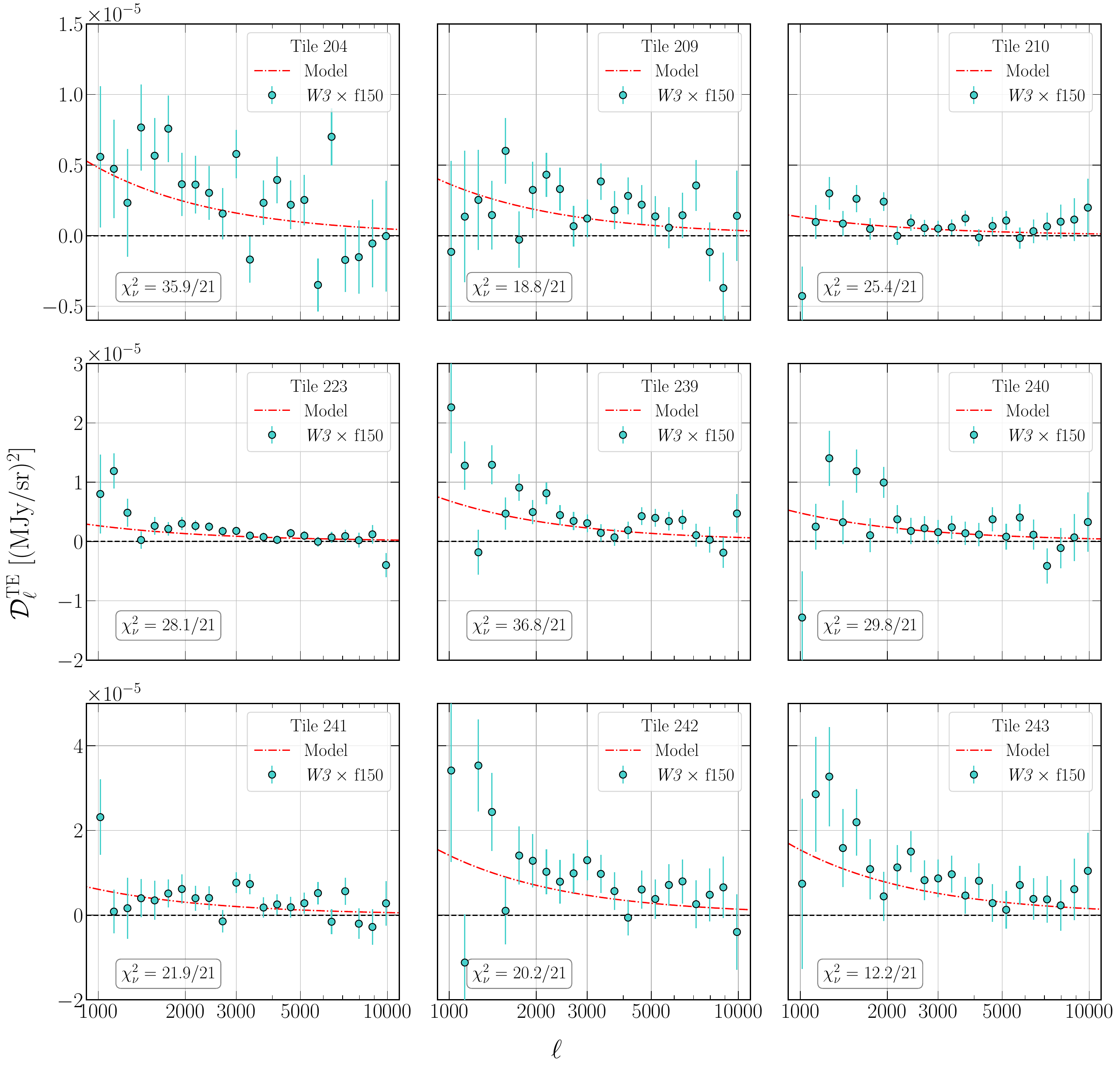}
    \caption{A selection of the highest signal-to-noise \textit{W3} $\times$ f150 $TE$ spectra. The best fit power law $C_\ell^{TE} \propto \ell^{-2.5}$ is shown (red dashed) along with its associated $\chi^2$ value.}
    \label{fig:3sigTE150}
\end{figure*}

\begin{figure*}
    \centering
    \includegraphics[width = \textwidth]{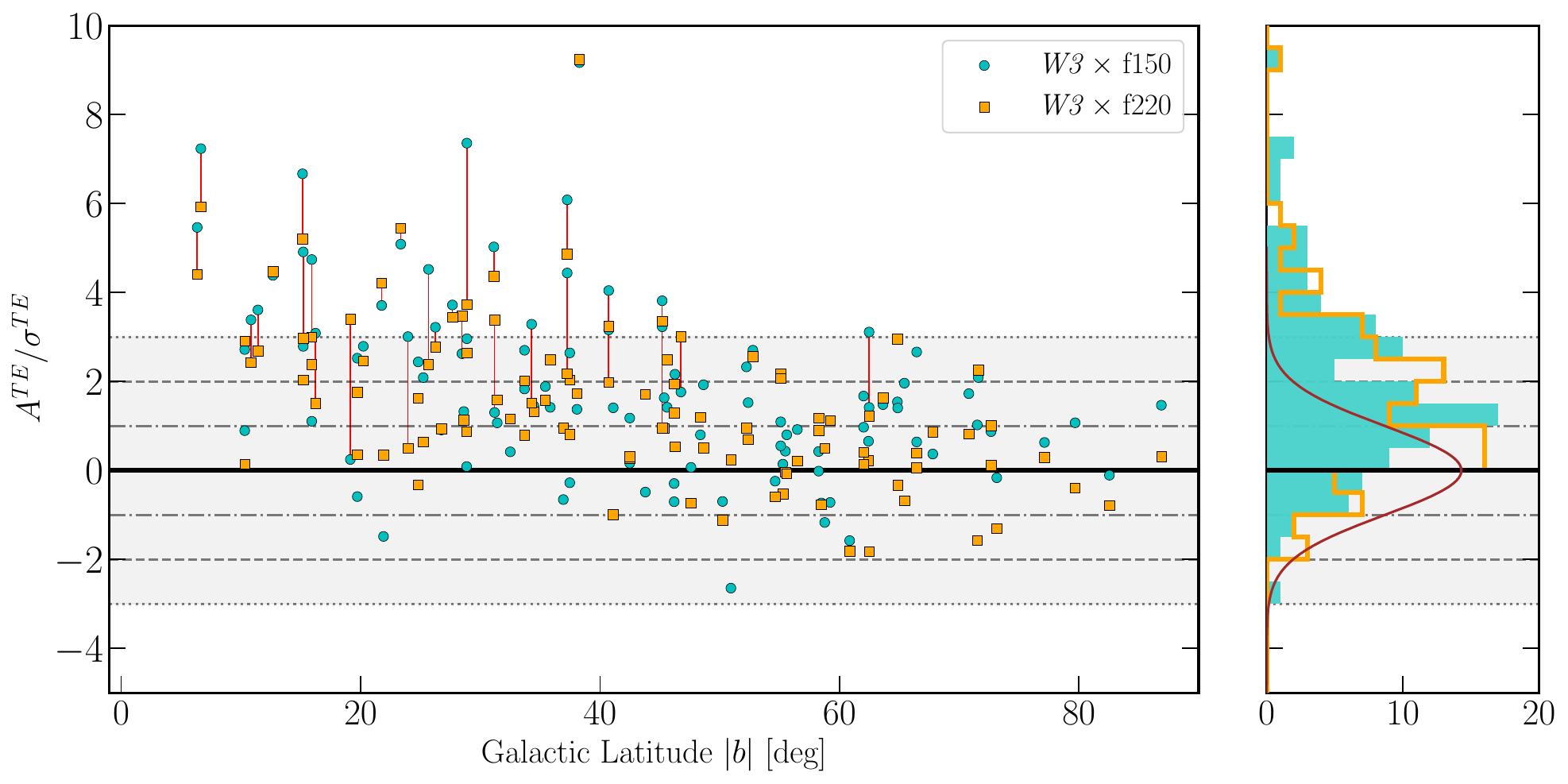}
    \caption{\textit{Left:} The best-fit $A_d^{TE}$ (see Equation~\eqref{eq:TEfit}) in each of the 107 regions for each of f150 and f220 is shown as a function of Galactic latitude. The $A_d^{TE}$ have been normalized by the uncertainty of the fit. Vertical red lines connect selected pairs of f150 and f220 values for the same region. \textit{Right:} Histogram of $A_d^{TE}/\sigma^{TE}$ with a unit Gaussian overlaid in dark brown, illustrating bias toward positive $TE$.}
    \label{fig:ensembleTE}
\end{figure*}

\begin{figure}
    \centering
    \includegraphics[width = \columnwidth]{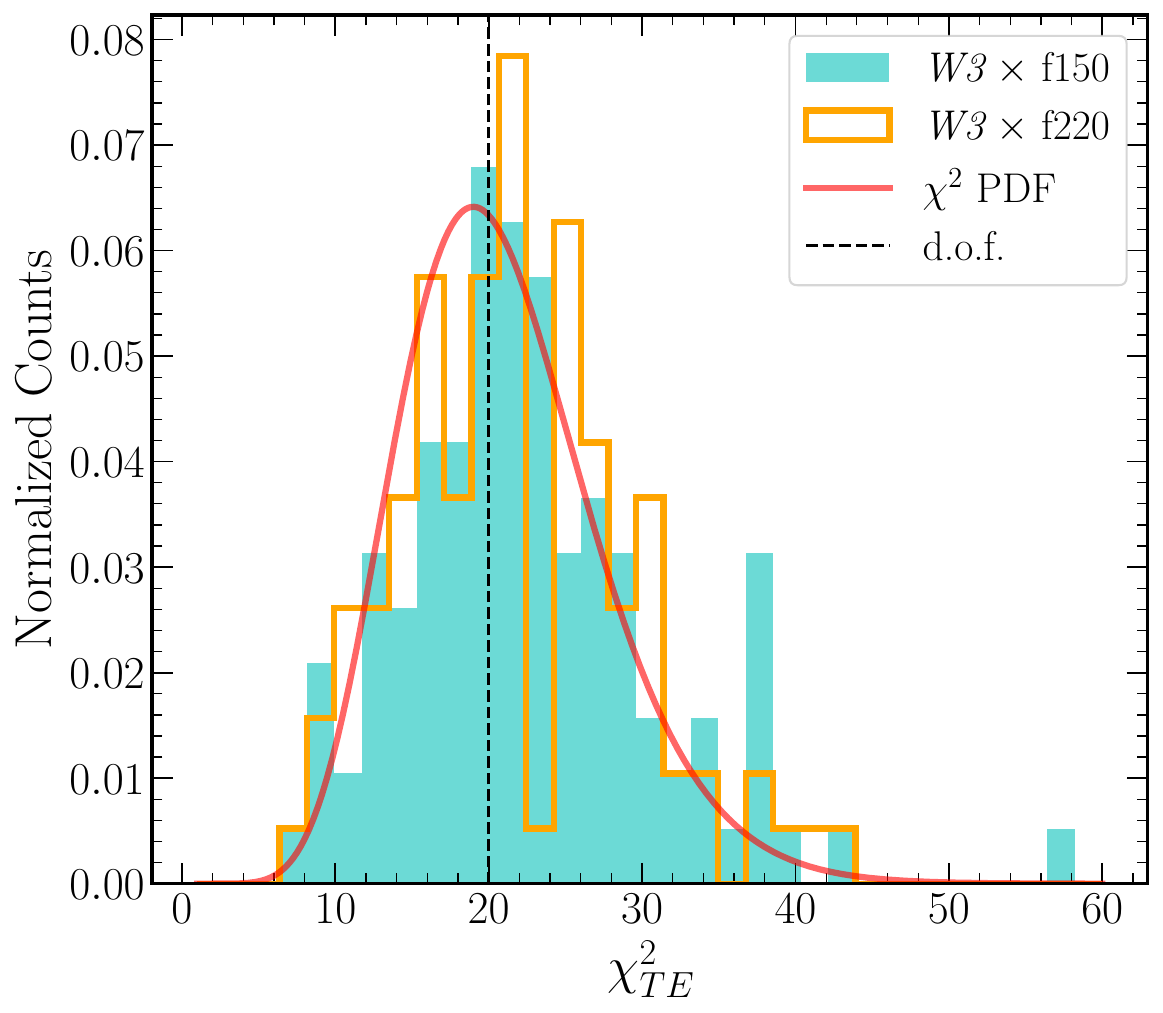}
    \caption{Distribution of $\chi^2$ values across all regions for the fits employing Equation~\eqref{eq:TEfit}. While a simple power law model $C_\ell^{TE} \propto \ell^{-2.5}$ is broadly consistent with the data, there is evidence for departures.}
    \label{fig:TE_chi2_comparison}
\end{figure}

\begin{figure}
    \centering
    \includegraphics[width = \columnwidth]{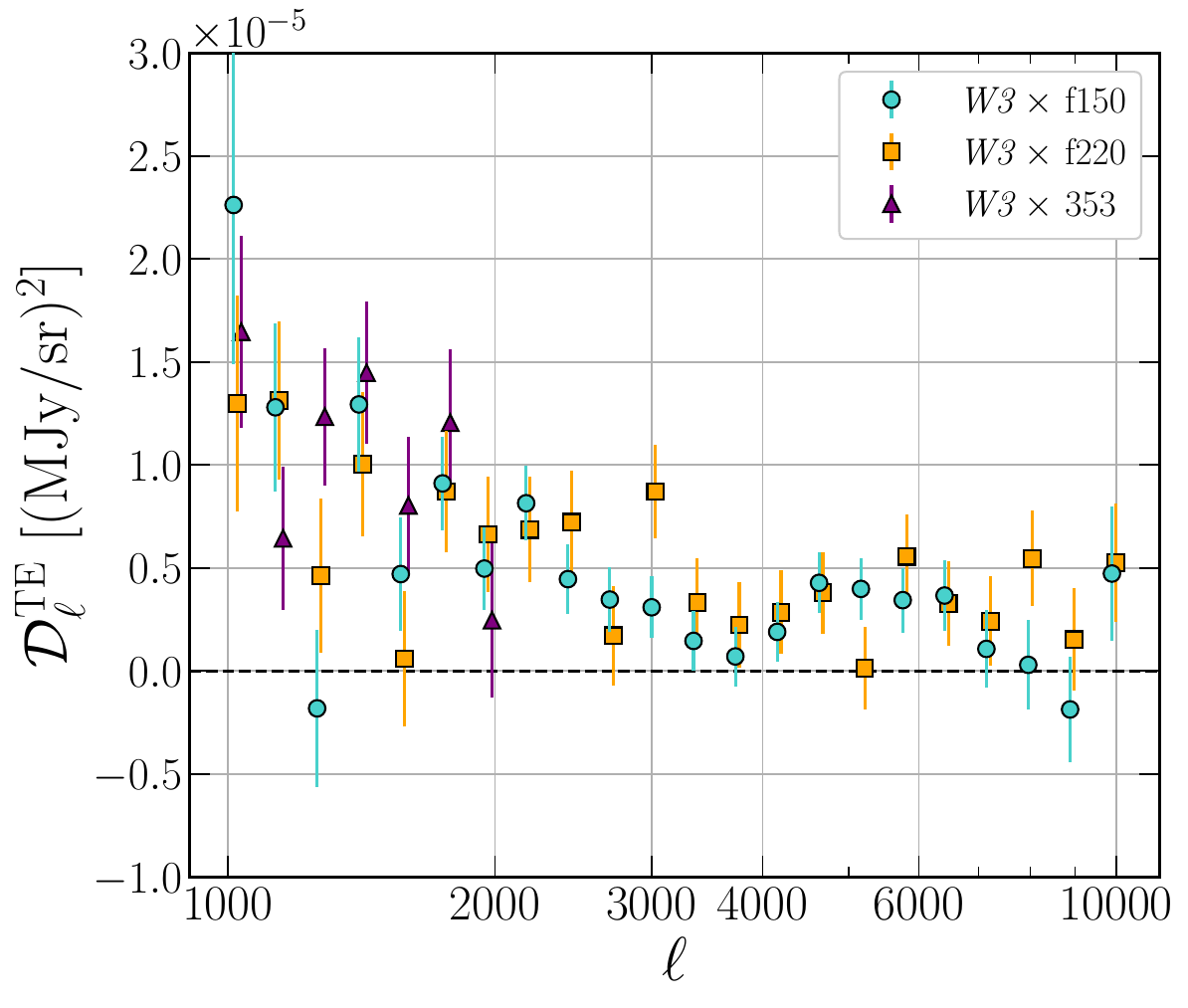}
    \caption{A comparison of the \textit{W3} $\times$ f150, f220, and Planck 353\,GHz $TE$ spectra of Tile~239 ((\textit{l, b}) =(172$^\circ$.2, -38$^\circ$.2)). The f220 and 353\,GHz spectra have been scaled to 150\,GHz assuming a modified blackbody spectral energy distribution with $\beta = 1.5$ and $T_d = 20$\,K, corresponding to multiplicative factors of 0.286 and 0.065, respectively. The Planck 353\,GHz spectrum is truncated at $\ell = 2000$ due to lack of Planck sensitivity at higher multipoles. The three spectra are broadly consistent in amplitude (after scaling) and shape over the full $\ell$ range shown.}
    \label{fig:239TE_3freq}
\end{figure}

\begin{figure}
    \centering
    \includegraphics[width = \columnwidth]{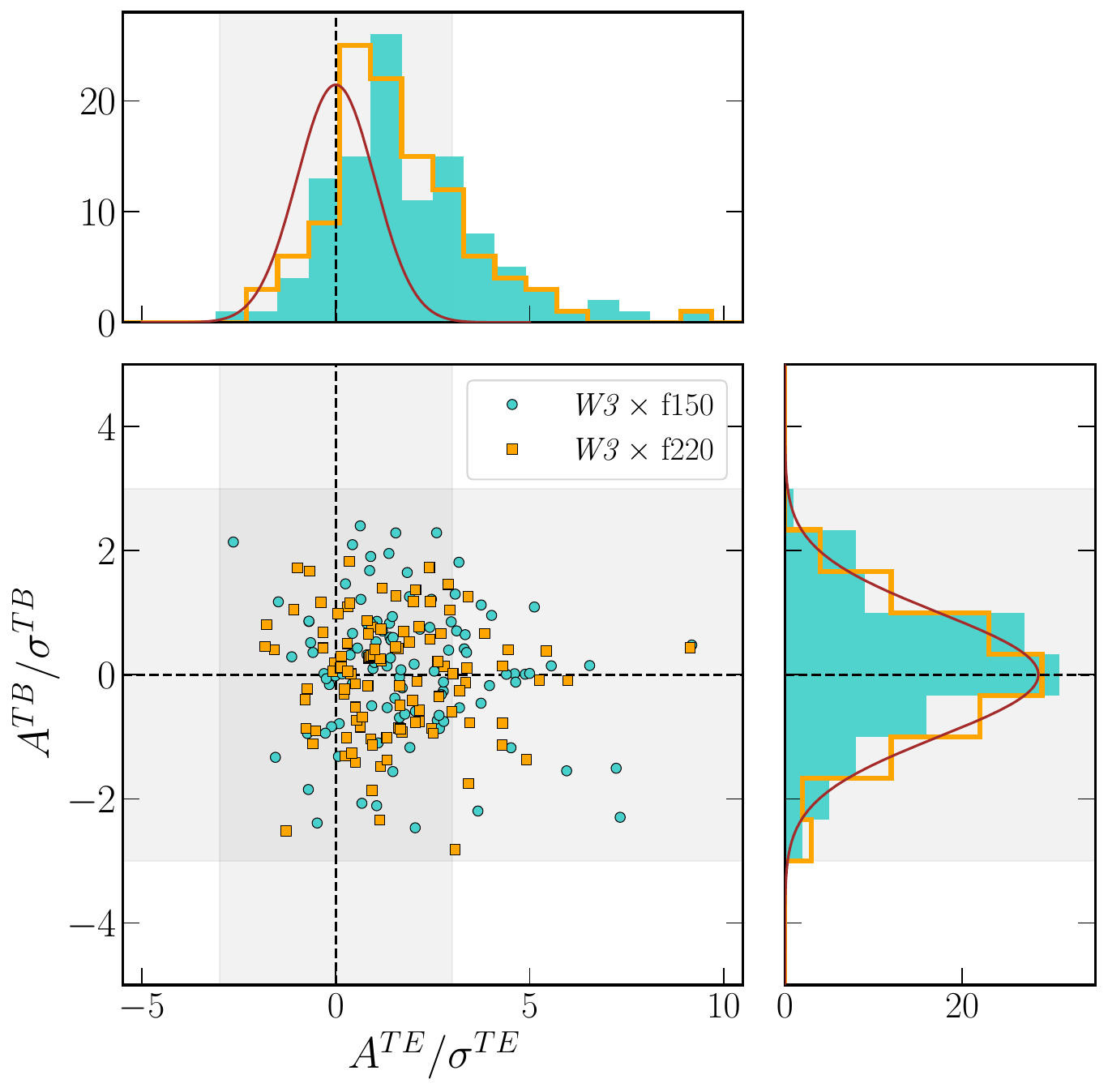}
    \caption{Best-fit $A_d$ values (see Equation~\eqref{eq:TEfit}) for $TE$ (x-axis) and $TB$ (y-axis) in each of the 107 regions at both f150 and f220. Both quantities have been normalized by the $1\sigma$ uncertainty of the fit with the shaded regions corresponding to $\pm3\sigma$. Also shown are the corresponding histograms with a unit Gaussian overlaid in dark brown. The distribution of $A^{TE}$ (top) is biased positive (as seen in Figure~\ref{fig:ensembleTE}), but the distribution of $A^{TB}$ (right) shows no significant trend.}
    \label{fig:TETB_histograms}
\end{figure}

In this Section, we analyze the cross power spectra between the {\it W3} total intensity map and millimeter polarization maps at 150, 220, and 353\,GHz, focusing on the $TE$ spectra. Unlike the $TT$ analysis in Section~\ref{sec:TT}, for the $TE$ analysis we employ ACT maps that have been co-added with Planck data \citep{2020JCAP...12..046N} to enhance signal-to-noise at $\ell \lesssim 2000$ (see Section~\ref{sec:ACT} for details).

We present a sample of nine of the highest signal-to-noise {\it W3}$\times{\rm f}150$ $TE$ spectra in Figure~\ref{fig:3sigTE150}. In all nine regions, $\mathcal{D}_\ell^{TE} > 0$ over most or all of the $\ell$ range considered ($10^3 < \ell < 10^4$). 

We first check if systematic temperature-to-polarization leakage in the ACT maps could bias our measurement of $TE$. Following \citet{lungu_2022}, the expected $T \rightarrow E$ leakage is modeled as a leakage beam $B^{T\rightarrow E}_{\ell}$ that is determined from ACT $Q$ and $U$ maps of Uranus. To first order, the expected bias to the $TE$ spectrum is given by $C_{\ell}^{TT} B^{T\rightarrow E}_{\ell}/B_{\ell}$, where $C_{\ell}^{TT}$ is the measured WISE $\times$ ACT cross spectrum and $B_{\ell}$ is the instrumental beam. The leakage signal is found to be largest in f220 and peaks at the highest multipoles included our analysis ($\ell \sim$4000), but is still $\lesssim 1\%$ of the measured $TE$ signal. We therefore ignore the systematic $T \rightarrow E$ leakage in the analysis.

We next apply the data model in Equation~\eqref{eq:TEfit} to characterize the spectra of all 107 regions. In our fiducial analysis, we use a uniform prior on $A_d$ and fix $\alpha_d^{TE} =-2.5$, representative of measurements over large sky areas at $40 < \ell < 600$  \citep{planck2016-l11A}. We fit over the range $10^3 < \ell < 10^4$ for both f150 and f220. We repeat the analysis with Planck 353\,GHz data, but restrict the fits to $1000 < \ell < 2000$ given the lack of constraining power of the Planck polarization data at higher multipoles.

The fits to the ${\it W3}\times{\rm f}150$ $TE$ spectra of the nine selected regions presented in Figure~\ref{fig:3sigTE150} demonstrate broad but imperfect agreement with a power law model. In all cases, $A_d^{TE} > 0$ with $A_d^{TE}/\sigma\left(A_d^{TE}\right) > 3$, indicating robust detections of a positive $TE$ signal in all nine regions. Figure~\ref{fig:ensembleTE} shows the $A_d^{TE}/\sigma\left(A_d^{TE}\right)$ values of all regions as a function of Galactic latitude. In total, there are 26 regions with $A_d^{TE}/\sigma\left(A_d^{TE}\right) > 3$ for f150, 17 regions for f220, and 13 regions with $A_d^{TE}/\sigma\left(A_d^{TE}\right) > 3$ in both f150 and f220. Most significant detections of non-zero $TE$ signal are in regions at $|b| < 40^\circ$ where the dust emission is brightest. There are no regions in which the $A_d^{TE} < 0$ at $3\sigma$ significance at either frequency. The distribution of fit $A_d^{TE}$ values is clearly biased toward $A_d^{TE} > 0$.

In detail, however, the $TE$ spectra are not all well-described by a power law in $\ell$. The $\chi^2$ distributions of the ensemble of f150 and f220 fits are presented in Figure~\ref{fig:TE_chi2_comparison}. Agreement with the expected $\chi^2$ probability density function (PDF) is generally good, but there are more regions with large $\chi^2$ values than expected from chance. Allowing $\alpha_d^{TE}$ to vary yields only marginal improvements in most regions, suggesting that the data have little constraining power on the $TE$ spectral index.

The shape of the $TE$ power spectrum may vary because of the structure of the diffuse dust emission itself, which is not necessarily well-described by a power law in $\ell$. On the other hand, the measured $TE$ power spectrum may be affected by any compact sources in the map that were not identified by the {\citetalias{{2014ApJ...781....5M}}} flags. Further, the WISE data still contain a number of unmitigated data artifacts that could also affect the spectra particularly at high $\ell$. Given the limitations imposed by these potential systematics and lack of evidence for variability in $\alpha_d^{TE}$, we do not draw strong conclusions on the shape of the $TE$ spectra.

Applying our fitting framework to the Planck 353\,GHz data yields 19 regions with $A_d^{TE} > 0$ at $3\sigma$ confidence. Five of these non-zero $TE$ detections are unique to the 353\,GHz analysis while the remaining regions were identified with f150 (11), f220 (11), or both (8). To assess the impact of the Planck data in the ACT+Planck co-adds, we repeat the f150 and f220 fits over the multipole range $2000 < \ell < 10^4$, where the Planck data have little constraining power. We find that the number of $3\sigma$ positive-$TE$ detections falls from 26 to 17 for f150 and from 17 to 5 for f220.

As further illustration of the relative constraining power of the three frequencies on the $TE$ spectrum, Figure~\ref{fig:239TE_3freq} presents the f150, f220, and Planck 353\,GHz $TE$ spectra of Tile~239, centered on (\textit{l, b}) =(172$^\circ$.2, -38$^\circ$.2). To facilitate direct comparison, we scale the f220 and 353\,GHz spectra to 150\,GHz assuming a modified blackbody emission law with $\beta = 1.5$ and $T_d = 20$\,K. The Planck 353\,GHz spectrum is truncated at $\ell = 2000$ due to lack of sensitivity at higher multipoles. There is broad agreement in the amplitude (after scaling) and shape of the $TE$ spectrum across the three frequencies. The ACT data are consistent with positive $C_\ell^{TE}$ out to scales $\ell \simeq$6000. To our knowledge, these are the smallest-scale measurements of a Galactic dust $TE$ spectrum to date. Using the \citet{Capitanio:2017} 3D reddening map, we estimate that the dust emission toward the center coordinates of Tile~239 primarily originates from a distance of about 150~pc from the Sun. At 150~pc, our measurements constrain the dust $TE$ down to physical scales of $\sim$0.05$~\mathrm{pc}$.

The fit $A_d^{TE}$ values are highly correlated with the fit $A_d^{TT}$ values, as expected. However, we find that the relation between these quantities is sub-linear---$A_d^{TE}$ scales roughly as $\left(A_d^{TT}\right)^{0.8}$. Two physical effects may contribute to this relation. First, as the dust column density increases, so too does the number of distinct structures that may be superimposed along the line of sight. While the $TT$ correlation is unaffected by such superposition, the $TE$ correlation may be weakened by depolarization and by loss of apparent filamentarity in the integrated map. Second, at higher column densities a loss of alignment between the local magnetic field and dust filaments is observed \citep{planck2014-XXXII}. This should weaken the $TE$ correlation, which is positive in diffuse regions due to a preferred alignment between dust structures and the magnetic field \citep{planck2015-XXXVIII, Clark:2021}.

We repeat this analysis on the $TB$ spectra, finding no 3$\sigma$ detections of non-zero $TB$ in any region. The fitted $A^{TB}$ amplitudes at both f150 and f220 are compared to the $A^{TE}$ amplitudes in Figure~\ref{fig:TETB_histograms}. Unlike the fit $A^{TE}$, the ensemble of $A^{TB}$ amplitudes show no bias toward positive or negative values at either frequency.

\section{Discussion}\label{sec:diss}

\subsection{Mid-Infrared PAH Emission as a Spatial Template for Microwave Dust Emission}
Measurements of the CMB at small ($\sim$arcminute) angular scales is a principal focus of current and upcoming ground-based experiments. Measurements of lensing of the CMB constrains the growth of structure in the Universe and the neutrino masses, while removal of the $B$-mode signature generated by lensing will be required for constraints on primordial $B$-mode signatures at the levels pursued by next-generation experiments. Critical to all of these analyses is robustness to Galactic emission at small angular scales. As the combination of sensitivity and angular resolution do not yet exist to characterize millimeter-wavelength dust emission at these scales directly, indirect probes of these dust properties are needed.

We have demonstrated that the WISE \textit{W3} measurements of Galactic dust emission correlate with both total and polarized intensity millimeter-wave observations at scales $10^3 < \ell < 10^4$. Further, the slope of the measured $TT$ spectrum is compatible with $TT$ spectra measured from millimeter data only at lower multipoles. The WISE data therefore offer a means of characterizing the spatial structure of Galactic dust emission, including non-Gaussianity, that will be informative for millimeter-wavelength analyses. 

While this is a promising direction for future analysis, we highlight a few caveats. \textit{W3} primarily traces emission from PAHs, which are known to differ in both emission physics and spatial distribution from the submicron grains responsible for the bulk of the millimeter emission. The fraction of the dust mass in PAHs is variable throughout the Galaxy, with fewer PAHs per unit dust mass found in very dense regions, in \ion{H}{2} regions, and in the Warm Neutral Medium (WNM). Because PAHs undergo single-photon heating, their emission scales with the first power of the energy density of the interstellar radiation field $U$. In contrast, millimeter-wavelength emission from submicron grains scales as $U^{1/4+\beta} \simeq U^{1/5.5}$. Thus, to the extent that the radiation field heating the dust varies across the sky, the ratio of PAH emission to millimeter dust emission likewise varies. On the other hand, the availability of two independent probes of $U$ could help constrain spatial variations in the dust temperature and thus frequency decorrelation, a key concern for $B$-mode analyses.

\subsection{Variability of the Dust Power Spectrum} \label{sec:discuss_var}
Simulated maps of Galactic emission used in cosmic microwave background analyses frequently assume that the Galactic dust $B$-mode spectrum is a power law in $\ell$ \citep[e.g.,][]{HerviasCaimapo2016, Thorne_2017}. Measurements with the Planck satellite demonstrate this to be a good approximation over large sky areas for $40 < \ell < 600$, with $C_\ell^{BB} \propto \ell^{-2.5}$ \citep{planck2016-l11A}. In detail, however, the dust power spectrum is expected to vary in slope across the sky. For instance, \citet{2021ApJ...908..186M} found that the steepness of the dust $TT$ spectrum is influenced by the filling factor of the WNM on the line of sight.

We have shown evidence of spatial variability in the steepness of the dust $TT$ spectrum at arcminute scales. Further, we find that some of the $TT$ spectra are not well-described by power laws at all. The spatial distribution of Galactic dust is complex, and it is unsurprising that a power law in $\ell$ is an inadequate description of the power spectrum. If this variability indeed arises from spatial inhomogeneity of interstellar density structures and the magnetohydrodynamic turbulence that shapes them, then corresponding variability is expected in polarization, including the dust $TE$ and $BB$ spectra. Detailed constraints on the shape of the dust $TE$ spectrum and its variations within the Galaxy will be possible with more sensitive polarization data.

The observed spatial variations in the dust power spectrum underscore the need for scrutiny of simple power law models for the scale dependence of Galactic dust emission. Moment-based methods that can account for deviations from power law behavior \citep[e.g.,][]{Chluba_2017,Azzoni_2023,Vacher_2023} are of particular interest and could be tested on the regions identified here.

A challenge for both the analysis presented in this work and for dust modeling in a CMB foregrounds context is the presence of compact sources. In addition to extragalactic sources, we have identified a number of Galactic sources such as planetary nebulae having strong emission at both mid-infrared and millimeter wavelengths. If left unmasked, these sources can strongly affect the measured power spectra at high $\ell$. While we have employed aggressive masking (as described in Section~\ref{sec:masking}), sources below the ACT flux density cut may still contribute non-negligible power. Dedicated identification and characterization of such sources will be the topic of future work.

\subsection{Dust TE and TB Correlations at Arcminute Scales}
Over large regions of sky and on large angular scales, the dust total intensity is positively correlated with the dust $E$-mode polarization \citep{planck2014-XXX, planck2016-l11A}. This positive $TE$ correlation is consistent with a preferential alignment between elongated dust intensity structures and the plane-of-sky projected magnetic field orientation traced by polarized dust emission \citep{planck2015-XXXVIII}. This magnetically-aligned density anisotropy is also seen in \ion{H}{1}, and additionally provides a natural explanation for the observation that the dust polarization $EE/BB > 1$ \citep{Clark_2015}. Filament-based models of Galactic dust polarization that invoke this alignment also show $TE > 0$ \citep{2019ApJ...887..136C, HerviasCaimapo+Huffenberger_2022}.  

If magnetically aligned ISM filaments source the observed positive $TE$ correlation, this raises several observationally measurable questions: in particular, whether and how $TE$ correlation changes as a function of scale and/or Galactic environment. There could be an environmental $TE$ dependence set by the relative orientation of filaments and magnetic fields in regions dominated by different physics. Filamentary structures are strongly aligned with the magnetic field orientation throughout the diffuse ISM \citep{Clark:2014}, but higher-density filaments are closer to being orthogonal to the projected magnetic field orientation \citep{planck2014-XXXII, planck2015-XXXV, Fissel:2019}. This empirical result may be related to the mass-to-flux ratio of molecular cloud filaments \citep[e.g.,][]{Seifried:2020}. 

A dust filament with a polarization structure that corresponds to a perpendicular plane-of-sky magnetic field orientation would produce a negative $TE$ correlation \citep{Zaldarriaga:2001, Huffenberger:2020}. We find that the measured $TE$ correlation is generally stronger at higher column densities and lower Galactic latitudes, where the dust is brighter and measured with higher signal-to-noise, but even at low Galactic latitudes we find no robust detections of negative $TE$ spectra (Figure \ref{fig:ensembleTE}). The data are thus consistent with a general alignment between the ACT-measured magnetic field and the density structures seen in PAH emission. The scale dependence of the $TE$ correlation is plausibly related to the physics that couples the dust density structure to the magnetic field on a particular scale. This work measures dust $TE$ that is generally biased toward positive values down to sub-parsec scales. 

Planck data also exhibit a non-zero $TB$ correlation over large sky areas and large angular scales \citep{planck2014-XXX, planck2016-l11A}. In the filament-based model, non-zero $TB$ is caused by imperfect alignment between the long axis of a dust filament and the magnetic field, such that non-zero $TB$ over large sky regions implies that this misalignment has a preferred handedness \citep{Huffenberger:2020, Clark:2021, Cukierman:2023}. In cross-correlation with the WISE data, we find no regions with robustly nonzero $TB$, and no preference for one sign of $TB$ over the distribution of sky regions considered.

\subsection{The Cosmic PAH Background} \label{sec:cib}
We have detected at $30\sigma$ significance a high-$\ell$ correlation between the {\it W3} map at 12\,$\mu$m and the f150 and f220 ACT maps inconsistent with extrapolation of the Galactic dust power spectrum. The correlation appears spatially isotropic and well-described as a power law $C_\ell \propto \ell^{-1}$. We conclude that this signal is extragalactic in origin and most likely to arise from the correlation between PAH emission in dusty, star-forming galaxies as seen by WISE and the CIB as seen by ACT.

The restframe MIR emission of a dusty star-forming galaxy is dominated by PAH features that can account for up to $\sim$20\% of its total infrared emission \citep{smith2007}. The strongest of these is the 7.7\,$\micron$ feature \citep{Tielens_2008}. Even at $z=0$ this feature makes a non-negligible contribution to the {\it W3} band, and it remains within the {\it W3} band up to $z \simeq$1. Using large optical galaxy catalogues, \citet{Chiang+Menard_2019} demonstrated that the \citetalias{2014ApJ...781....5M} map is correlated with galaxies in redshift bins up to $z \sim 2$, consistent with redshifted PAH emission. Detailed modeling of the extragalactic background light suggests that the 12\,$\micron$ extragalactic sky is dominated by PAH emission from star-forming galaxies rather than by emission associated with active galactic nuclei \citep[AGN, e.g.,][]{2018MNRAS.474..898A}. Likewise, in recent multi-wavelength fits to galaxies detected by ACT, \citet{Kilerci_2023} found that even galaxies dominated by AGN emission at ACT frequencies could be dominated by PAH emission at 12\,$\micron$. Thus, the ``Cosmic PAH Background'' appears the most natural explanation for the observed correlation, though we cannot rule out significant contributions from a $12\,\micron$ ``Cosmic AGN Background'' on the basis of these data alone.

The cross-power spectrum encodes the relationship between the galaxies producing the MIR and the millimeter wavelength emission. We find the correlation is inconsistent with a pure Poisson spectrum, but, as illustrated in Figure~\ref{fig:CIB_150MCMC}, the relative contributions of a clustered versus Poisson component is not well constrained. Interpretation of shape of the power spectrum would benefit from forward models of the Cosmic PAH Background based on galaxy simulations.

The frequency spectrum of the emission, both in the MIR and at millimeter wavelengths, is a window into galaxy properties. Similarly, quantification of the level of correlation of maps of diffuse extragalactic emission at two frequencies constrains the diversity of emission spectra and their variability with galaxy properties and with cosmic time. For instance, with a greater number of MIR bands, it will be possible to assess whether different PAH features preferentially arise from galaxies with different properties. Likewise, the level of correlation between MIR and millimeter wavelength maps constrains the extent to which it is the same population of galaxies responsible for the observed emission in both frequency ranges. Ultimately, the implementation of PAH emission spectra in tools such as SIDES \citep{bethermin2017} and Websky \citep{stein2020} could allow these data to place constraints on the relationship between PAH-bright galaxies and those responsible for the CIB, including how the PAH luminosity function evolves with cosmic time.

In addition to extragalactic emission, the observed high-$\ell$ correlation could also include a contribution from Galactic point sources. Large numbers of dusty, compact Galactic point sources have been identified in Planck data \citep{planck2014-a37}, and ACT has observed objects such as planetary nebulae that are also bright in the {\it W3} band \citep{2020JCAP...12..046N}. While we see no evident correlation between the high-$\ell$ component we model as extragalactic emission and Galactic latitude or dust column density, careful treatment of Galactic contamination will be required to make quantitative comparisons between the signal observed here and models of extragalactic emission.

\section{Conclusions}\label{sec:conclu}
We have presented a correlation analysis between 12\,$\mu$m emission observed by WISE and both 150 and 220\,GHz emission observed by ACT at multipoles $10^3 < \ell < 10^4$. Our principal conclusions are as follows:

\begin{itemize}
    \item We report a $30\sigma$ detection of a spatially isotropic, high-$\ell$ $TT$ signal that we interpret as a correlation between the CIB at ACT frequencies and the ``Cosmic PAH Background'' seen by WISE. The spectrum is well-fit by a power law $C_\ell \propto \ell^{-1}$, consistent with a clustered component. The fits do not require, but do not exclude, the presence of a Poisson component.
    \item The $TT$ spectrum of Galactic dust at $10^3 < \ell < 10^4$ is generally well-fit by a power law in $\ell$ with $C_\ell^{TT} \propto \ell^{-2.9}$, consistent with Galactic dust $TT$ spectra that have been measured at lower multipoles \citep[e.g.,][]{Gautier_1992, Bracco:2011, 2012ApJ...744...40H, planck2013-pip56}. However, we find evidence for spatial variability in the power law index and identify several regions where a power law is an inadequate description of the $TT$ spectrum at the sensitivity of the measurements. The strength of the observed correlation suggests that WISE maps of dust emission can be used to understand the spatial statistics of millimeter-wavelength dust emission at small angular scales.
    \item We identify 35 regions with $>3\sigma$ detections of positive $TE$ correlation and none with $>3\sigma$ detections of negative $TE$ correlation. We further find that the distribution of all fit $TE$ amplitudes is biased positive. To our knowledge, these are the highest-$\ell$ measurements of the dust $TE$ correlation to date.
\end{itemize}

This work showcases the power of high angular resolution observations of dust emission at MIR wavelengths to understand the astrophysics of dust emission at millimeter wavelengths. The small-scale dust morphology in the WISE maps may be representative of what will be observed by next-generation millimeter experiments, and so characterization of its non-Gaussianity and other properties is a promising direction for future work. 

The Cosmic PAH background provides another window into the evolution of galaxies with cosmic time, especially the buildup of PAHs. Implementation of PAH spectra into existing tools to model the CIB and cross-correlating the Cosmic PAH Background with other tracers of galaxy properties (e.g., \ion{H}{1} emission) will be important for understanding the properties of the galaxies giving rise this emission. Given the recent JWST detection of the 2175\,\AA\ feature, associated with PAHs, in a $z = 6.71$ galaxy \citep{Witstok_2023}, understanding how the Universe becomes enriched with PAHs is all the more pressing.

The data underlying the analyses in this work are set to improve dramatically in the near future. The Simons Observatory will soon begin operations in Chile and will provide maps of millimeter dust emission and polarization with greater sensitivity and over a wider frequency range than ACT \citep{SO_Science, Hensley_2022}. The SPHEREx satellite will soon measure the full sky at $\sim$6$\arcsec$ angular resolution in 102 channels spanning 0.75--5\,$\mu$m, each with comparable sensitivity to WISE \citep{Crill_2020}. From the Cosmic PAH Background to the morphology of dust throughout the Milky Way, these new datasets can be used to extend the investigations presented here.

\section*{Acknowledgments}

We thank B. Draine, D. Finkbeiner, J. Greene, A. Goulding, A. Meisner, M.A. Miville-Deschênes, and D. Spergel for helpful conversations and guidance throughout the course of this work. Support for ACT was through the U.S. National Science Foundation through awards AST-0408698, AST- 0965625, and AST-1440226 for the ACT project, as well as awards PHY-0355328, PHY-0855887 and PHY-1214379. Funding was also provided by Princeton University, the University of Pennsylvania, and a Canada Foundation for Innovation (CFI) award to UBC. ACT operated in the Parque Astronómico Atacama in northern Chile under the auspices of the Agencia Nacional de Investigacion y Desarrollo (ANID). The development of multichroic detectors and lenses was supported by NASA grants NNX13AE56G and NNX14AB58G. Detector research at NIST was supported by the NIST Innovations in Measurement Science program.

We thank the Republic of Chile for hosting ACT in the northern Atacama, and the local indigenous Licanantay communities whom we follow in observing and learning from the night sky. 

Computing was performed using the Princeton Research Computing resources at Princeton University. R.C.R. acknowledges support from the Ford Foundation Predoctoral Fellowship from the National Academy of Sciences, Engineering, and Medicine. S.E.C. acknowledges support from the National Science Foundation grant No. AST-2106607. S.K.C. acknowledges support from NSF award AST-2001866. C.S. acknowledges support from the Agencia Nacional de Investigaci\'on y Desarrollo (ANID) through FONDECYT grant no.\ 11191125 and BASAL project FB210003.

\facilities{Planck, WISE}

\software{Astropy \citep{astropy:2013, astropy:2018}, \texttt{emcee} \citep{emcee, emcee_v3}, Matplotlib \citep{Matplotlib}, \texttt{NaMaster} \citep{2019MNRAS.484.4127A}, NumPy \citep{NumPy}, \texttt{pixell} \citep{2021ascl.soft02003N}, SciPy \citep{SciPy}}

\bibliographystyle{aasjournal}
\bibliography{references, Planck_bib}

\end{document}